\documentclass[a4paper,12pt]{article} 

\usepackage{epsfig} 
\usepackage{amssymb} 
\usepackage[tbtags]{amsmath}
\usepackage{upref} 
\usepackage[normal]{caption}
\usepackage{float}
\usepackage{wrapfig}
\usepackage{subfigure}
\usepackage{longtable}
\def\asusy{$a_{\mu}^{{\rm \small SUSY}}$}
\def\bsg{$b\to s\gamma$}
\def\bmumu{B_s^0\to\mu^+\mu^-}

\def\higgsu{m_{H_2}^2}
\def\higgsd{m_{H_1}^2}

\def\neumass{m_{\tilde\chi_1^0}}

\def\vev#1{\langle#1\rangle}
\def\higgsu{m_{H_u}^2}
\def\higgsd{m_{H_d}^2}

\def\neumass{m_{\tilde\chi_1^0}}

\newcommand{\crosssec}{\sigma_{\tilde\chi^0_1-p}}
\def\tanb{\tan\beta}

\def\neut{\tilde\chi_1^0}
\def\bsg{$b\to s\gamma$}

\def\asusy{a^{\rm SUSY}_\mu}

\def\relic{\Omega_{\tilde{\chi}_1^0}}

\newcommand{\captions}{\sf\caption}
\def\fig#1{Fig.\,\ref{#1}}
\def\eq#1{(\ref{#1})}

\def\lsim{\raise0.3ex\hbox{$\;<$\kern-0.75em\raise-1.1ex\hbox{$\sim\;$}}}
\def\gsim{\raise0.3ex\hbox{$\;>$\kern-0.75em\raise-1.1ex\hbox{$\sim\;$}}}

\begin{document}

\pagestyle{empty}

\rightline{UdeM-GPP-TH-05-132}
\rightline{IPPP/05/10}
\rightline{DCPT/05/20}
\rightline{KIAS-P05030}
\rightline{KAIST-TH 2005/07}
\rightline{FTUAM 05/06}
\rightline{IFT-UAM/CSIC-05-21}
\rightline{hep-ph/0505019}
\rightline{May 2005}

\renewcommand{\thefootnote}{\alph{footnote}}

\vspace{0.cm}
\begin{center}
  {\large{\bf
      %% TITLE %%
      Direct detection of neutralino dark matter in supergravity
      \\[5mm]
      }}
  \vspace{0.5cm}
  \mbox{
    \large{
      %% AUTHORS %%
      S.~Baek\,$^{1}$,
      D.G.~Cerde\~no\,$^{2}$
      Y.G.~Kim\,$^{3}$,
      P.~Ko\,$^{4}$, and
      C.~Mu\~noz\,$^{5,\,6}$
      }
    }
  \vspace{0.2cm}
  
  {\small
    {\it 
      %% AFFILIATIONS %%
      ${^1}$
      %School of Physics, KIAS, Seoul 130-722, Korea\\
      Laboratoire Ren\'e J.-A. L\'evesque, Universit\'e de Montr\'eal, 
      C.P. 6128, succ. centre-ville, Montr\'eal, QC, Canada H3C 3J7\\
      \vspace*{2mm}
      ${^2}$ 
      Institute for Particle Physics Phenomenology, University of
      Durham, DH1 3LE, UK\\
      \vspace*{2mm}
      ${^3}$ 
      Department of Physics, Korea University, Seoul 136-701, Korea\\
      \vspace*{2mm}
      ${^4}$
      School of Physics, KIAS, Seoul 130-722, Korea\\
      \vspace*{2mm}
      ${^5}$ 
      Department of Physics, KAIST, Daejon 305-701, Korea \\
      \vspace*{2mm}
      ${^6}$ 
      Departamento de F\'{\i}sica
      Te\'orica C-XI and Instituto de F\'{\i}sica
      Te\'orica C-XVI,\\ 
      Universidad Aut\'onoma de Madrid,
      Cantoblanco, 28049 Madrid, Spain.
      } 
    }
  
  \vspace{1cm}
  
  {\bf Abstract} 
  \\[7mm]
  \begin{minipage}[h]{14.0cm}
    %% ABSTRACT %%
The direct detection of neutralino dark matter is analysed in general
supergravity scenarios, where non-universal soft scalar and gaugino
masses can be present. In particular, the theoretical predictions for the
neutralino-nucleon cross section are studied and compared with the
sensitivity of dark matter detectors. We take into account the most
recent astrophysical and 
experimental constraints on the parameter space, including the 
current limit on B$(\bmumu)$. 
The latter puts severe limitations on the dark matter scattering cross
section, ruling out most of the regions that would be within the reach
of present experiments. 
We show how 
this constraint can be softened with the help of appropriate
choices of non-universal parameters which increase the Higgsino
composition of the lightest neutralino and minimise the chargino
contribution to the $b\to s$ transition. 
  \end{minipage}
  \end{center}
\newpage

\setcounter{page}{1}
\pagestyle{plain}
\renewcommand{\thefootnote}{\arabic{footnote}}
\setcounter{footnote}{0}

\section{Introduction}
\label{intro}

A long-lived or stable weakly interacting massive particle (WIMP) is
a particularly attractive 
candidate for dark matter in the Universe \cite{mireview}. The interest
in WIMPs resides mainly 
on the fact that they can be present in the right
amount to account for the matter density observed in the analysis of
galactic rotation curves \cite{Persic}, cluster of
galaxies
and large scale flows \cite{Freedman}, 
$0.1\lsim \Omega\, h^2\lsim 0.3$ 
($0.094\lsim\Omega\, h^2\lsim 0.129$
if we take into account the recent data obtained by the
WMAP satellite \cite{wmap03-1}).

Many underground experiments are being carried out around the world in
order to detect the flux of WIMPs on the Earth, by observing their
elastic scattering on target nuclei through nuclear recoils
\cite{mireview}.  Although
one of the current experiments, the DAMA collaboration \cite{dama},
has 
reported data favouring the existence of a WIMP signal
with a WIMP-proton cross section $\approx 10^{-6}-10^{-5}$ pb 
for a WIMP mass smaller than $500-900$ GeV \cite{dama,halo}, 
other collaborations such as 
CDMS Soudan \cite{experimento2}, EDELWEISS \cite{edelweiss}, 
and ZEPLIN I \cite{zeplin1} 
claim to have excluded important regions of the DAMA 
parameter space. 
In the light of these experimental results more than 20 experiments
are
running or in preparations around the world.
For example, this is the case of GEDEON \cite{IGEX3}, which will be
able to explore positively a WIMP-nucleon cross section $\sigma \gsim
3\times 10^{-8}$ pb. CDMS Soudan will be able to test in the future
$\sigma\approx 2\times 10^{-8}$ pb, and 
the very sensitive detector
GENIUS \cite{HDMS2} will test a WIMP-nucleon cross section
$\sigma\approx10^{-9}$ pb. In fact, already 
planned detectors working with 1 tonne of Ge/Xe
\cite{xenon} are expected to reach cross sections as low as
$10^{-10}$ pb.

The leading candidate for WIMP is 
the lightest neutralino, $\neut$,
a particle predicted by the
supersymmetric (SUSY) extension of the standard model. 
Given the experimental situation, and assuming that the dark matter 
is a neutralino, it is natural to wonder how big 
the cross section for its direct detection can be.
Obviously,
this analysis is crucial in order to know the
possibility of detecting dark matter 
in the experiments.
In fact, the analysis of the neutralino-proton cross section 
has been carried out by many authors and during many 
years \cite{mireview}.
The most recent studies take into account the present
experimental and astrophysical constraints
on the parameter space. 
Concerning the former, 
the lower bound on the Higgs mass,
the $b\to s\gamma$ branching ratio, and the
muon anomalous magnetic moment, $a_\mu\equiv(g_\mu-2)/2$, 
have been considered.
The astrophysical bounds on the dark matter density,
have also been
imposed on the theoretical computation of the relic neutralino
density, assuming thermal production.
In addition, 
the constraints that the absence of dangerous charge
and colour breaking minima imposes on the parameter space
have also been taken into account \cite{cggm03-1}.

Recently, the importance of the improved experimental upper limit on 
the $\bmumu$ branching ratio \cite{bmumuexp,bmumuexp2}
was stressed in Ref.\,\cite{bkk04-1}, 
where a strong correlation was found between this
observable and the spin-independent neutralino-nucleon cross section, 
the origin of which resides in $\tan\beta$ and 
the neutral Higgs boson masses  $(m_H,\,m_A)$.  
In particular, both observables increase for large $\tan\beta$ and low
values of the Higgs masses. 
For this reason, some of
regions where $\crosssec$ can be consistent with the future dark
matter detectors are excluded.

We will work within the context of supergravity theories, taking the
soft super\-symmetry-breaking parameters of the MSSM 
as inputs at the Grand Unification scale,
$M_{GUT}\approx 2\times 10^{16}$ GeV, and solving the renormalization
group equations (RGEs) to obtain the supersymmetric spectrum at the
electroweak scale.
In the particular case of the minimal supergravity scenario (mSUGRA),
where the soft terms 
are considered to be universal
at the GUT scale, the predictions for B$(\bmumu)$ are currently 
below the reach
of Tevatron \cite{adkt02-1,bkk04-1,eos05-1}. 
Therefore no further constraint on the theoretical predictions for the
neutralino-nucleon cross section appears and the usual upper limit,
$\crosssec\lsim 10^{-8}$ pb, is obtained.

Relaxing the universality condition is a more generic situation within
the framework of supergravity. The presence of non-universal soft
scalar
\cite{Fornengonew,Arnanew,Bottino,arna2,Arnowitt,Santoso,Drees,Nojiri,
darkcairo,Arnowitt3,nosopro,Dutta,Rosz,Profumo,cggm03-1,Farrill} 
and gaugino masses
\cite{Nath2,darkcairo,nosopro,Orloff,Dutta,Birkedal,Roy2,cggm03-1}
has been extensively considered in the literature. Non-universalities
in both the scalar and gaugino sectors have also been studied
\cite{pallis,cm04-1}.
Certain choices of non-universalities in these 
scenarios were shown to lead to a sizable increase of the theoretical
predictions for $\crosssec$.
Nevertheless, in such cases the
constraint on B$(\bmumu)$ plays a very important role and large
regions in the parameter space are forbidden \cite{bkk04-1} (see also
\cite{Rosz} for an analysis in a model with minimal $SO_{10}$ SUSY breaking).

For example, 
non-universal scalar masses can lead to 
an enhancement in
$\crosssec$, mainly due to the associated
reduction of the Higgs masses. 
The most important effect is
due to non-universal soft Higgs masses, which can  be parameterised 
at the GUT scale as
follows:
\begin{equation}
  m_{H_{d}}^2=m^{2}(1+\delta_{1})\ , \quad m_{H_{u}}^{2}=m^{2}
  (1+ \delta_{2})\ .
  \label{Higgsespara}
\end{equation}
The optimal choice to
increase the neutralino detection cross section is $\delta_1<0$ and
$\delta_2>0$. 
Nevertheless, the decrease in the Higgs masses also leads to large
values of B$(\bmumu)$, which can be in conflict with the experimental
constraint. 
As a consequence, those regions of the parameter space where the
neutralino scattering is within the reach of the DAMA experiment are
ruled out. 
Actually, in some cases, 
the upper bound on the 
B$(\bmumu)$ branching ratio becomes even stronger than the upper bound 
from CDMS experiment \cite{bkk04-1}.

On the other hand, non-universal gaugino masses allow more flexibility
in the neutralino sector. 
The following parameterisation can be used
\begin{eqnarray}
  M_1=M\ , \quad M_2=M(1+ \delta'_{2})\ ,
  \quad M_3=M(1+ \delta'_{3})
  \ ,
  \label{gauginospara}
\end{eqnarray}
where $M_{1,2,3}$ are the bino, wino and gluino masses, respectively,
and $\delta'_i=0$ corresponds to the universal case. 
A decrease in $M_3$ can induce both a decrease in the Higgs mass,
through its effect on the RGE of
$m_{H_u}^2$, and an increase of the Higgsino components of $\neut$.
Despite the associated enhancement of $\crosssec$,
the current limit on
B$(\bmumu)$ puts a strong constraint in the large $\tan\beta$ region.

In this paper we analyse the most general situation, where both
scalar and gaugino masses are allowed to be non-universal at
$M_{GUT}$. This is an interesting possibility, since neutralinos whose
detection cross section can be within the reach of future experiments
can appear with a wide range of masses, from over 400 GeV to almost 10
GeV \cite{cm04-1}.
As we will see, the new experimental result for B$(\bmumu)$ has an
important effect.

\section{Non-universal scalar and gaugino masses}
\label{main}

In our analysis the soft supersymmetry-breaking terms, parameterised
according to 
\eq{Higgsespara} and \eq{gauginospara}, 
are taken as inputs 
at the high energy scale
$M_{GUT}$,
where unification of the 
gauge coupling constants takes place.
In addition, the ratio of the Higgs vacuum
expectation values, $\tan\beta\equiv\vev{H_u^0}/\vev{H_d^0}$ is a free
parameter, as well as 
the sign of the Higgsino mass parameter, $\mu$,
which remains 
undetermined by the minimisation of the Higgs potential.

The most recent experimental and astrophysical
constraints will be taken into account. In particular, the lower
bounds on the masses of the supersymmetric particles and on the
lightest Higgs have been implemented, as well as the experimental
bounds on the branching ratio of the \bsg\ process 
($2\times10^{-4}\le$B(\bsg))$\le 4.1\times10^{-4}$ from the
measurements of $B\to X_s\gamma$ decays at CLEO \cite{cleo} and BELLE
\cite{belle})
and on 
$\asusy$.  The evaluation of the neutralino relic density is carried
out with the program {\tt micrOMEGAs} \cite{micro}, and, due to its
relevance, the effect of the WMAP constraint on it will be shown
explicitly.
Dangerous charge and colour breaking minima of the Higgs potential
will be avoided by excluding unbounded from below (UFB) 
directions. Finally, the improved
experimental constraint on the $\bmumu$ branching ratio,
B$(\bmumu)<2.9\times10^{-7}$, obtained from a combination of the
results of CDF
\cite{bmumuexp} and D0,
\cite{bmumuexp2}, will be included.

Concerning $\asusy$, we have taken into account the
recent experimental result for the muon
anomalous magnetic moment \cite{g-2}, as well as the most recent
theoretical evaluations of the Standard Model contributions
\cite{newg2,hagiwara,troconiz}. It is found that when $e^+e^-$ data
are used the experimental excess in $a_\mu$
would constrain a
possible supersymmetric contribution to be
$\asusy=(27.1\,\pm\,10)\times10^{-10}$ \cite{newg2}. 
However, 
when tau data are used a smaller
discrepancy with the above experimental measurement is found.
Due to this reason, in our analysis we will not impose this
constraint, but only indicate the regions compatible with it
at $2\sigma$ level, this is, 
$7.1\times10^{-10}\le\asusy\le47.1\times10^{-10}$.

Even after including the tau data, positive values of $\asusy$ are
clearly 
favoured. Because of this and the fact that the sign of $\asusy$ is
basically given by $\mu M_2$, we will only consider the case where
$sign(M_2)=sign(\mu)$\footnote{
 A different sign for $M_2$ and $\mu$
 could in principle also be used, thus obtaining $a_{\mu}^{{\rm \small
 SUSY}}<0$.  
 Nevertheless, since the negative values of $\asusy$ which are 
 allowed by tau data are small in modulus, very large values of
 $|M_2|$ are necessary. Therefore this possibility is very constrained.}. 
Similarly, the constraint on the \bsg\
branching ratio is much weaker when $sign(M_3)=sign(\mu)$. 
Finally, variations in the sign of $M_1$ do not induce significant
changes in the allowed regions of the parameter space (e.g., its
effect on $\asusy$, due to diagrams with neutralino intermediate states,
is smaller than the one of $M_2$). However, when
$sign(M_1)=sign(\mu)$ the theoretical predictions for $\crosssec$ are
larger. 
For these reasons we will restrict our analysis to positive values of
$M_{1,2,3}$ and $\mu>0$. 
Note in this sense, that due to the
symmetry of the RGEs, the results for ($M_{1,2,3},\mu,A$) are
identical to those for ($-M_{1,2,3},-\mu,-A$).

We will be mostly interested in analysing the conditions under which
high values for the cross section are obtained. For this reason, we
will concentrate on some interesting choices for scalar
non-universalities, 
exemplified by the following cases \cite{cggm03-1}
\begin{eqnarray}
  &&a)\quad\delta_{1}=0,\quad\ \  \delta_2=1;\nonumber\\ 
  &&b)\quad\delta_{1}=-1,\quad \delta_2=0;\nonumber\\ 
  &&c)\quad\delta_{1}=-1,\quad \delta_2=1,
  \label{nunivhiggs}
\end{eqnarray}
and study the effect of adding gaugino non-universalities to these.

For illustrative purposes, let us first briefly review the case with
only non-universal Higgs masses.

For low values of $\tan\beta$ ($\tan\beta\lsim20$) the most important
constraint on the 
parameter space is that on the lightest Higgs mass, which excludes
regions with low values for the common gaugino mass, $M$. This sets
an upper bound for the neutralino-nucleon cross section 
such that compatibility
with the present dark matter experiments is not achieved.
In these cases
the region excluded due to the constraint on B($\bmumu$)
is small and contained within the area excluded by the Higgs constraint.
Therefore, it
does not lead to a further restriction in the allowed area.
It is worth reminding that thanks to the Higgs non-universality the
UFB constraints are more easily fulfilled than in mSUGRA \cite{cggm03-1}.

The theoretical predictions for the
neutralino-nucleon cross section increase for larger values of
$\tan\beta$. However, at the same time the experimental
constraints on
\bsg\ and $\bmumu$ become more stringent\footnote{
  Notice, however, that when a general flavour mixing among squarks is
  taken into account, a calculation of B($\bmumu$) beyond the leading
  order in the large $\tan\beta$
  regime may show a reduction with respect to its value under the
  assumption of
  minimal flavour violation \cite{okumura}.
} 
and can exclude those regions
with larger values of $\crosssec$ \cite{bkk04-1}.

\begin{figure}[!t]
  \hspace*{-1.5cm}\epsfig{file=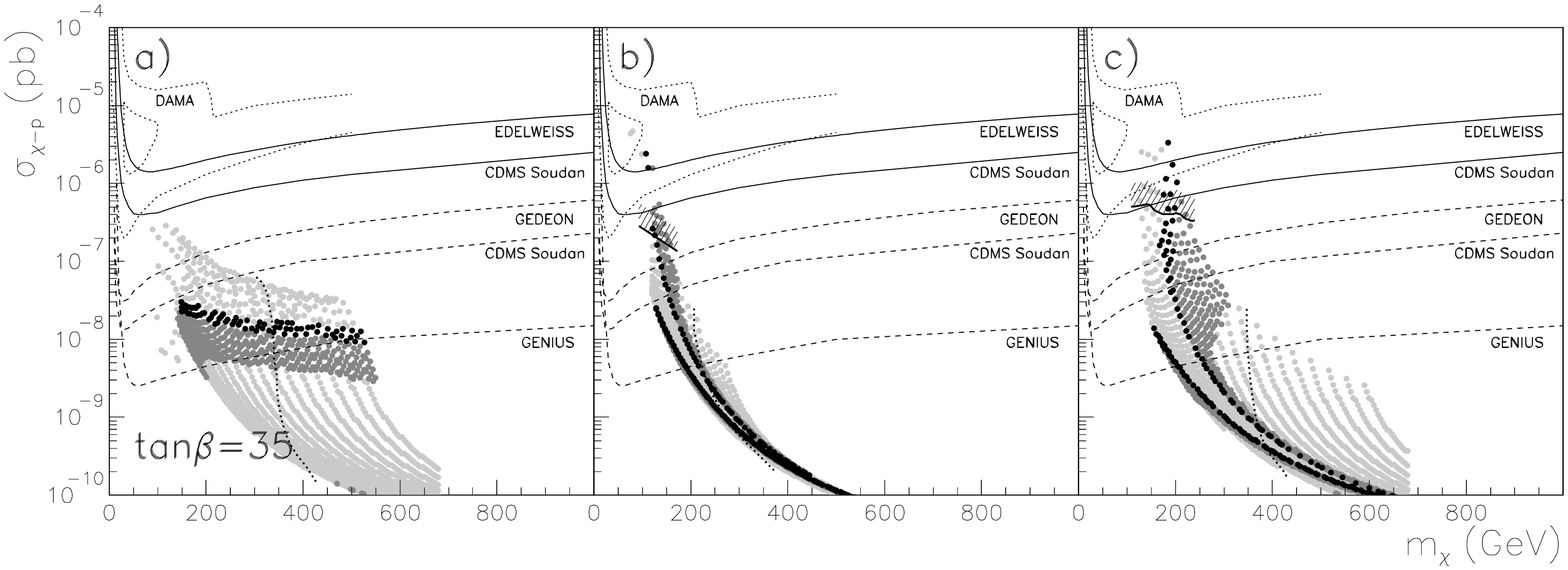,width=18cm}
  \captions{Scatter plot of the
    scalar neutralino-proton cross section $\crosssec$ as a function
    of the neutralino mass $\neumass$ for universal gauginos,
    $\delta'_{2,3}=0$, and the three choices for
    non-universal scalars \eq{nunivhiggs} in a case with
    $\tan\beta=35$
    and $A=0$. The light grey dots correspond to points fulfilling
    the experimental and UFB 
    constraints. The dark grey dots represent points
    fulfilling in addition 
    $0.1\le\relic\,h^2\le0.3$ and the black ones correspond to those
    consistent with
    the WMAP range. 
    The area to the right of the dotted line is disfavoured by
    $g_{\mu}-2$ if $e^+e^-$ data is used.  
    Points above the solid line with upper shading are excluded by the
    experimental constraint on B($\bmumu$).
    The sensitivities of present and projected dark matter 
    experiments are
    also depicted with solid and dashed lines, respectively.
    The large (small) area bounded by dotted lines is allowed by the
    DAMA experiment when astrophysical uncertainties are (are not)
    taken into account.
  }
  \label{nunivsc35}
\end{figure}

Let us illustrate this with an example.
The theoretical predictions for the
neutralino-nucleon cross section 
as a function of the neutralino mass 
are represented in Fig.\,\ref{nunivsc35} for a scan in the parameter
space where $m\le1500$ GeV and $50$ GeV$\le M\le1500$ GeV,
with
$\tan\beta=35$, $A=0$, and the three choices of non-universal Higgs
masses \eq{nunivhiggs}. 
Light grey dots in the figure represent points fulfilling
all the experimental and UFB constraints. Dark grey dots correspond to
those 
points where the neutralino relic density falls in the range
$0.1\le\relic\,h^2\le0.3$, 
and those points consistent with the WMAP range are
represented in black. 
The region to the right of the dotted line does not fulfil the
constraint on $\asusy$ when $e^+e^-$ data are used.
Due to the importance of the constraint on B($\bmumu$), its effect 
is shown explicitly in
the figure by means of a solid line with upper shading. 
Points above that line
are excluded for having a too large value of B($\bmumu$).

There are obvious differences in the predicted $\crosssec$
obtained in case a)
and in cases b) and c). Whereas in case a) the
non-universal structure of the Higgs parameters is such that the $\mu$
term is particularly reduced, 
in cases b) and c) it is the decrease in $m_A$ which 
is more important, owing to
the decrease of $\higgsd$ at the GUT scale.
This entails a larger increase of $\crosssec$ in these last cases, but
also leads to an enhanced B($\bmumu$) which can be in conflict with
the experimental bound. As a consequence of this, all the points with  
$\crosssec\gsim10^{-6}$ pb are ruled out. Part of the remaining points
are within the reach of the projected GEDEON detector. 
In case a) $\crosssec\lsim3\times10^{-8}$ pb and would
escape detection in the projected CDMS Soudan experiment and could
only be tested in the GENIUS detector.

\begin{figure}[!t]
  \hspace*{-1.5cm}\epsfig{file=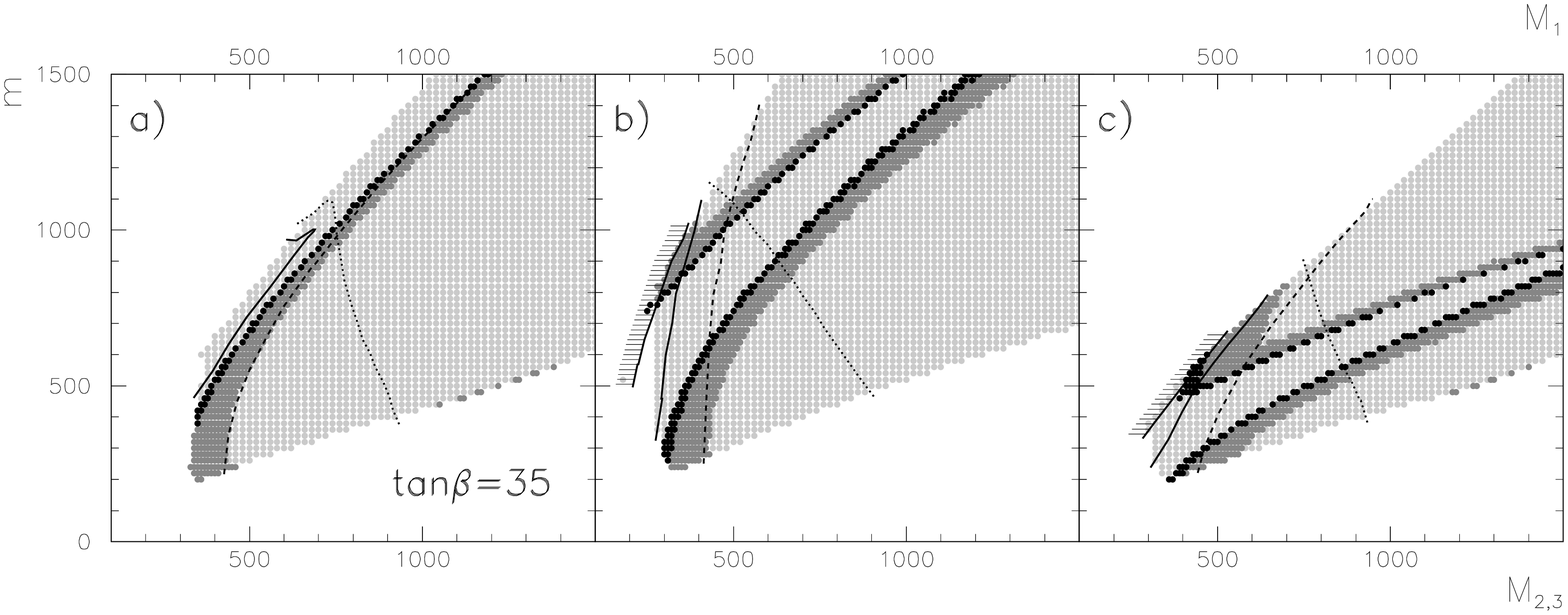,width=18cm}
  \captions{Effect of the different experimental constraints on
    the parameter space $(m,M_i)$ for universal gauginos,
    $\delta'_{2,3}=0$, and the three choices for
    non-universal scalars \eq{nunivhiggs} in a case with
    $\tan\beta=35$
    and $A=0$.
    The region to the right of the dotted line is disfavoured by
    $g_{\mu}-2$ if $e^+e^-$ data is used. 
    The area to the left of 
    the solid line with upper shading is ruled out by the
    experimental constraint on B($\bmumu$).
    The light shaded area is favoured by the remaining
    experimental and UFB constraints, while the dark one fulfils in
    addition $0.1\leq \Omega_{\tilde{\chi}_1^0}h^2\leq 0.3$. The black
    region on top of this indicates the WMAP range, 
    $0.094\leq \Omega_{\tilde{\chi}_1^0}h^2\leq 0.129$.
    The zone to the left of the thick solid line is within the
    projected sensitivity of the CDMS Soudan
    experiment. Similarly, the region to the left of the thick dashed
    line corresponds to those points that will be accessible to the
    GENIUS experiment. 
  }
  \label{nunivssparam35}
\end{figure}

To understand the effect of the different constraints it is 
illustrative to represent the excluded regions in the $(m,M)$
parameter space. This is done in Fig.\,\ref{nunivssparam35}, where
the same colour
convention as in Fig.\,\ref{nunivsc35} is used for denoting points
with different values of the relic density. Points to the
left of the solid line with upper shading are excluded by the
bound on B($\bmumu$), and those to the upper right of the dotted line
are disfavoured by $g_\mu-2$ when $e^+e^-$ data are used.
The regions compatible with the sensitivities of
the projected
CDMS Soudan and GENIUS detectors correspond to those on the 
left of the solid and dashed lines, respectively.

In case a) all the points within the reach of the projected CDMS
Soudan detector lie in the region where $\relic\,h^2<0.094$, in the
upper left corner. However, most of the points in agreement with the
WMAP result (black dots) 
are within the reach of GENIUS. Interestingly, this occurs
for a wide range of values of $m$ and $M$.
Contrariwise, in cases b) and c) CDMS Soudan can
test some of the points in agreement with WMAP. 
This occurs when the CP-odd Higgs mass is
sufficiently small,
in the proximity of the region which is 
ruled out for having $m_A^2<0$, in the upper
left corner of the plot. Notice that this is the case for $M\lsim250$
GeV, and a part of 
those points is disallowed by the constraint on B($\bmumu$). In these
two cases, points with $M\lsim550$ GeV could be tested by GENIUS.

\begin{figure}[!t]
  \hspace*{-1.5cm}\epsfig{file=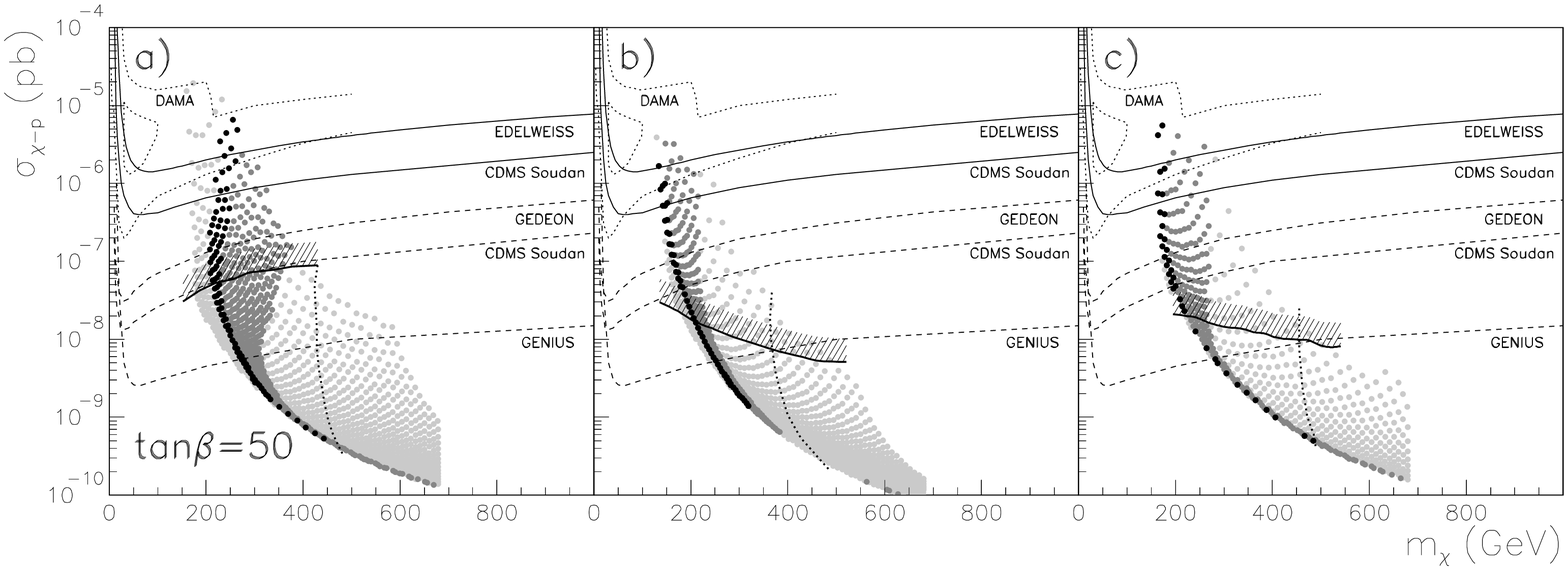,width=18cm}
  \captions{The same as Fig.\,\ref{nunivsc35}, but for
  $\tan\beta=50$.}
  \label{nunivsc50}

  \hspace*{-1.5cm}\epsfig{file=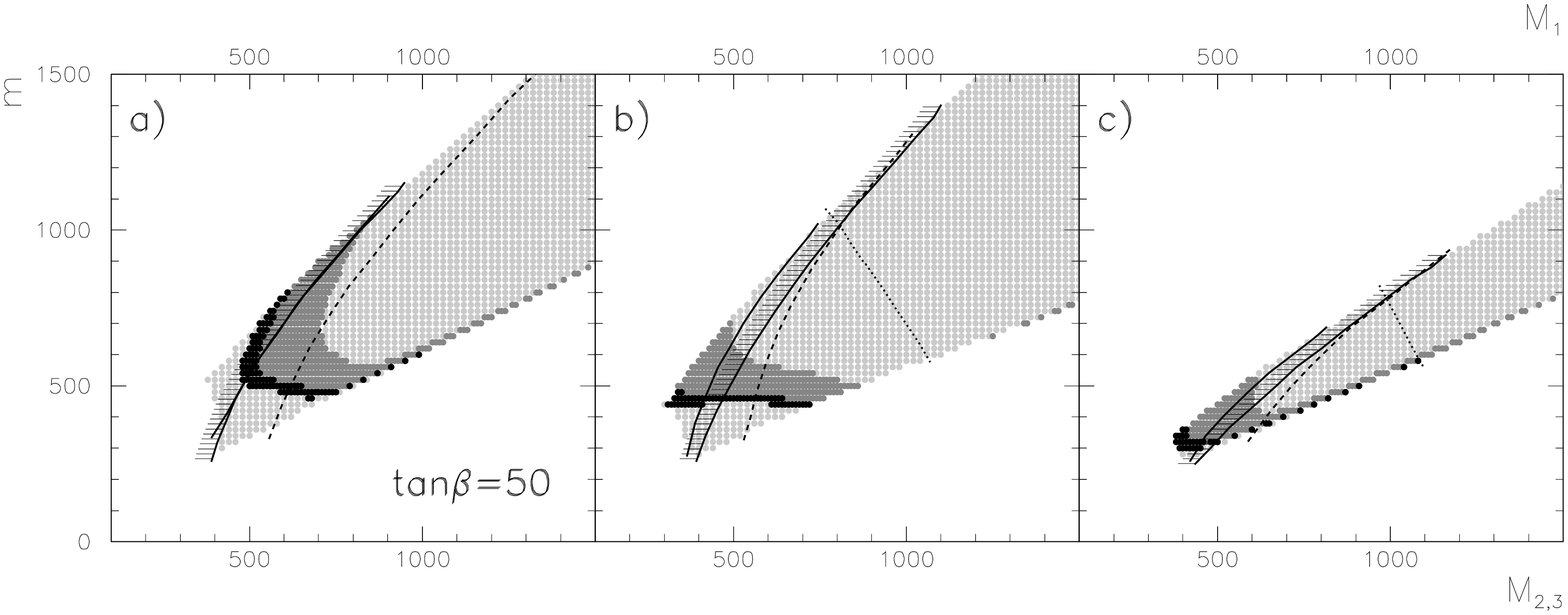,width=18cm}
  \captions{The same as Fig.\,\ref{nunivssparam35}, but for
$\tan\beta=50$. 
  }
  \label{nunivssparam50}
\end{figure}

With larger values of $\tan\beta$ the resulting $\crosssec$ could be
even within the reach of present dark matter detectors. 
Let us study this possibility and concentrate on the case 
$\tan\beta=50$. The corresponding theoretical predictions for
$\crosssec$ are represented in Fig.\,\ref{nunivsc50}. 
Although in principle, regions of the parameter space are found where
the neutralino detection cross section can be as large as
$\crosssec\gsim10^{-6}$ pb while having the correct relic density,
the associated increase in B($\bmumu$) is such that the experimental
constraint is violated and extensive
areas are ruled out. Owing to this, 
points compatible with present detector sensitivities disappear and
an upper bound of $\crosssec\lsim10^{-7}$ pb is obtained.

The corresponding $(m,M)$ parameter space is plotted in 
Fig.\,\ref{nunivssparam50}. For such high values of
$\tan\beta$ the three choices of non-universal Higgs masses,
a), b), and c) allow a large reduction in
the CP-odd Higgs mass. 
In particular, this reduction is larger than the decrease of $\mu$ in
case a) and for this reason the three examples present a large
resemblance. 
Once more the regions which are within the
reach of dark matter detectors are very disfavoured by the predicted
values of B($\bmumu$). As we can see, in the three cases 
the areas excluded by this
constraint enclose all the points within the reach of the projected 
CDMS Soudan. GENIUS would be able to test some of the remaining points
which have the correct value for the relic density. Note that most
of the points of the parameter space which could escape detection at
GENIUS are located along the coannihilation tail, where the neutralino
and the light stau are almost degenerate in mass.

At this point it may seem that the observed 
correlation between B($\bmumu$) and $\crosssec$ is inevitable and
that therefore large neutralino detection cross sections, within the
reach of present experiments, are not attainable. However, 
this correlation can
be diluted under several circumstances. For instance, 
the gluino mediated contribution to the $b\to s$ transition can have
the opposite sign than
the chargino mediated term, which is typically dominant,
thereby leading to a partial cancellation and slightly decreasing 
B($\bmumu$).

For a larger reduction, one can 
consider tuning the value of the top trilinear coupling,
$A_t$, at the GUT scale in such a way that the stop ($\tilde t_L -
\tilde t_R$) mixing is reduced and the stop mass increased.
Consequently, the chargino
mediated $b\to s$ transition 
is suppressed. This can be done with $A_t>0$ at the GUT scale.
For large values of $\tan\beta$, for which the $\mu/\tan\beta$ term in
the stop mixing can be neglected, the chargino contribution to
B($\bmumu$) can be qualitatively expressed as
\begin{equation}
{\rm B}(\bmumu)\propto
  \frac{\tan^6\beta}{m_A^4}\left(\frac{\mu A_t}{m_{\tilde
  t_L}^2}\right)^2\ .
  \label{bmumu_approx}
\end{equation}
When larger and positive values for 
$A_t$ are taken at the GUT scale, its value at the electroweak scale,
after applying the RGEs, becomes less negative. Thus $A_t^2$
decreases, $m_{\tilde t_{L,R}}^2$ increase through the effect of $A_t$ on
their RGEs, and 
as a consequence, the term in parenthesis in \eq{bmumu_approx} becomes
smaller.
Such a modification of $A_t$ also causes a decrease in the lightest
Higgs mass. This, together with the enhancement of the Higgsino
components of $\neut$, is helpful for obtaining an increase in
$\crosssec$ 
but one has to make sure the experimental bound on $m_h$ is not
violated. 

\begin{figure}[!t]
  \epsfig{file=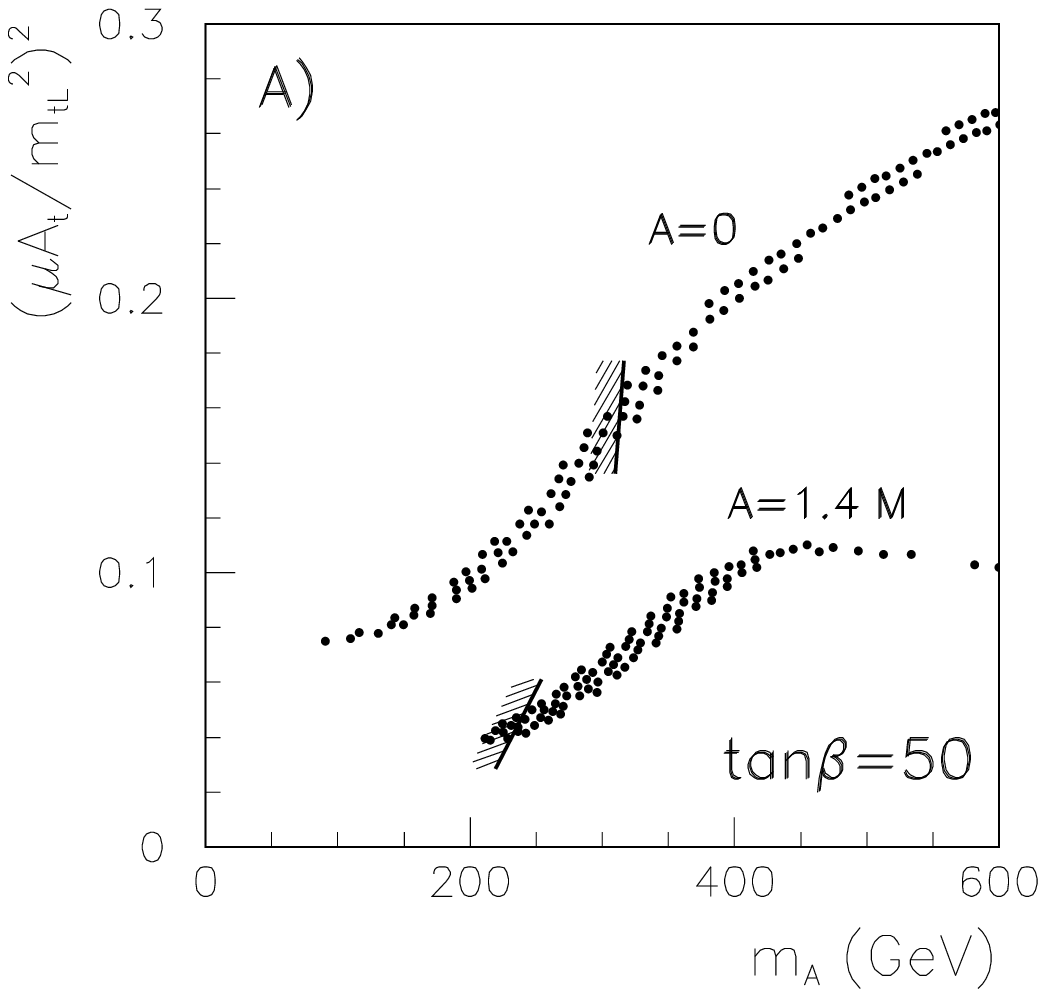,width=8cm}\hspace*{-0.5cm}
  \epsfig{file=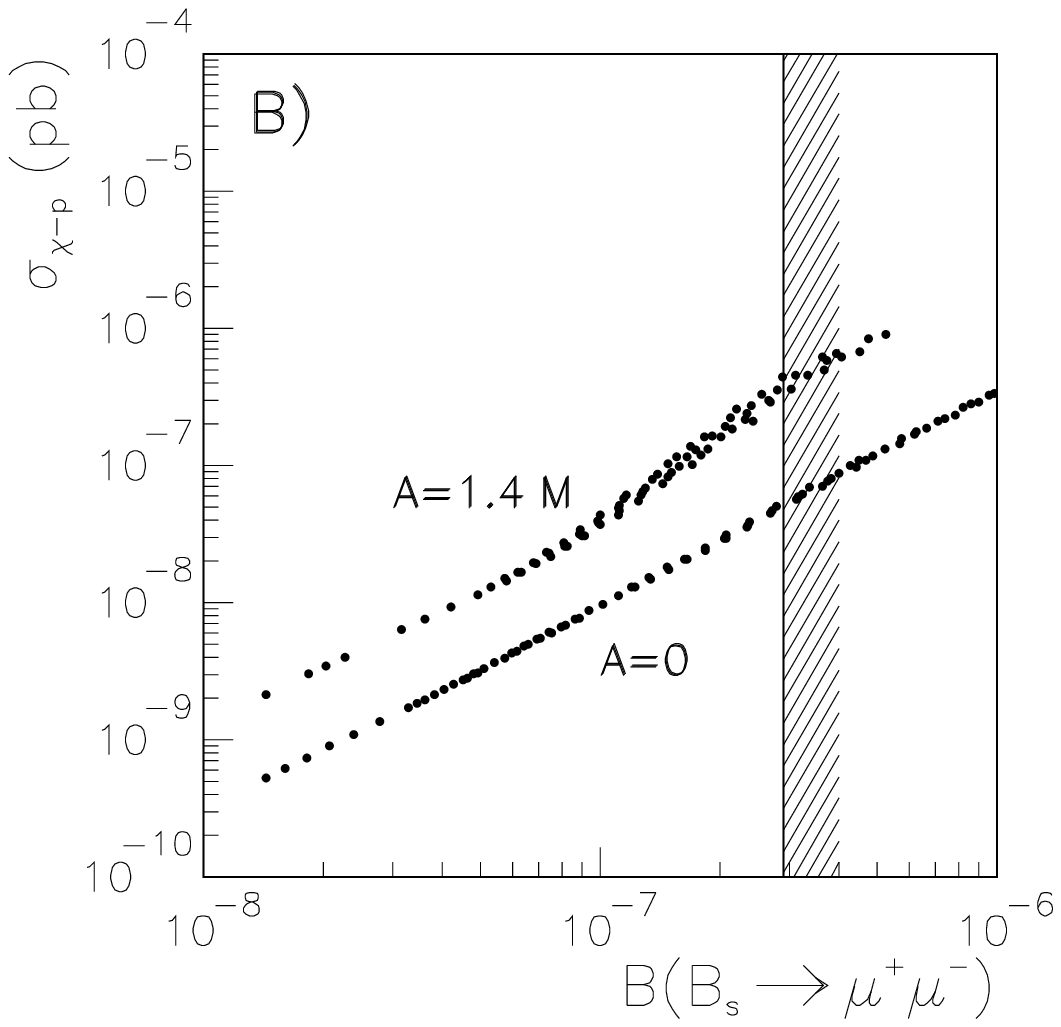,width=8cm}
  \captions{A) Ratio $(\mu A_t/m_{\tilde t_L}^2)^2$ 
  versus the CP-odd Higgs 
  mass for two cases with
  $\tan\beta=50$,
  non-universal Higgses with $\delta_1=0$ and $\delta_2=1$ in
  \eq{Higgsespara}, and
  $A=0,\, 1.4M$. Only those points fulfilling all the experimental
  constraints and whose relic density is in agreement with WMAP are
  plotted. The effect of the upper limit on B($\bmumu$) is shown
  explicitly by means of a solid line with shading, which
  excludes the points to its left. B) Associated theoretical
  predictions for $\crosssec$ versus B($\bmumu$). The experimental
  limit on B($\bmumu$) is represented by the vertical line with right
  shading. 
  }
  \label{alla}
\end{figure}

\begin{figure}[!t]
  \begin{center}
  \epsfig{file=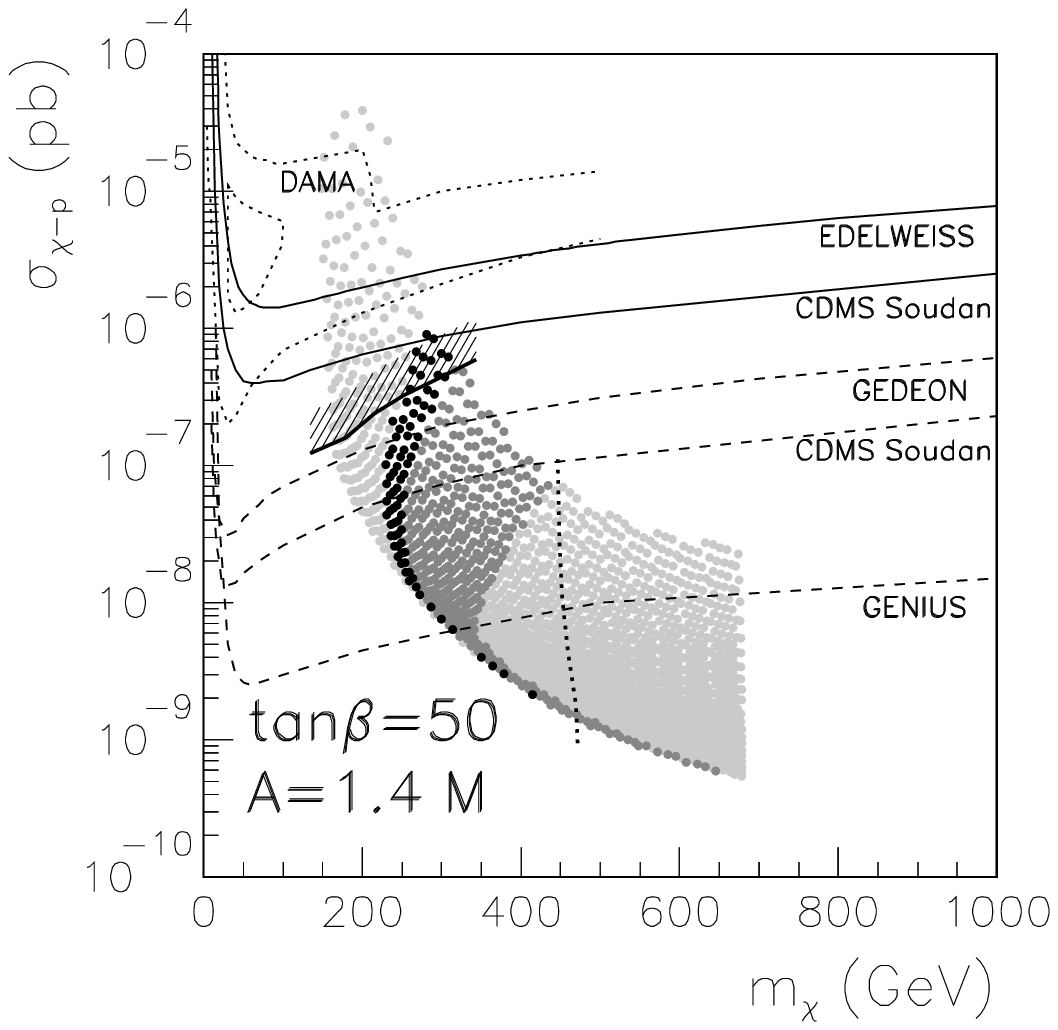,width=8cm}
  \end{center}
  \captions{The same as Fig.\,\ref{nunivsc35}, but for a case with
  $\tan\beta=50$, non-universal Higgses with
  $\delta_1=0$ and $\delta_2=1$ in 
  \eq{Higgsespara}, and $A=1.4M$.}
  \label{a14m}
\end{figure}

In order to exemplify this behaviour 
let us concentrate on the case with
$\tan\beta=50$ and non-universal Higgs masses according to case a) in
\eq{nunivhiggs}, and consider variations in the trilinear
parameter. 
For example, let us compare the case where $A=0$, which was
already shown in \fig{nunivsc35}, with the one where 
$A=1.4\,M$. The ratio $(\mu A_t/m_{\tilde t_L}^2)^2$
is represented on the left hand side of \fig{alla}
as a function of the  CP-odd Higgs mass for both cases, where we
have scanned in the whole $(m,M)$ parameter space, and applied all the
experimental and astrophysical constraints.
As we can
see, $(\mu A_t/m_{\tilde t_L}^2)^2$ is considerably smaller 
for the example with $A=1.4\,M$ and
for this reason the predictions for B($\bmumu$) decrease. As a consequence,
smaller values of the CP-odd Higgs mass are viable. Owing to this and
to the enhancement of the Higgsino composition of the lightest
neutralino that originates from the decrease of $\mu$, the theoretical
predictions for $\crosssec$ become larger.
This is illustrated on the
right-hand side of \fig{alla}, where 
$\crosssec$ is represented versus
B($\bmumu$) for both $A=0$ and $A=1.4\, M$.
An increase of one order of magnitude in $\crosssec$ is observed for
$A=1.4\,M$ and the resulting detection cross section,
$\crosssec\lsim4\times10^{-7}$ pb is close to the present CDMS Soudan
sensitivity. 
Finally, the resulting neutralino detection cross section as a
function of the neutralino mass is represented in \fig{a14m}, which
can be compared with \fig{nunivsc50}a, and evidences the less strict
bound on B($\bmumu$).
For $A\gsim1.4\,M$ the ratio $(\mu A_t/m_{\tilde t_L}^2)^2$ does not
decrease further and consequently the maximum value of $\crosssec$ 
does not
continue increasing.
Needless to say, 
the optimal choice of $A$ that 
leads to maximal $\crosssec$ 
is very dependent on the rest of the initial parameters.

The decrease in the ratio $(\mu A_t/m_{\tilde t_L}^2)^2$, and therefore in
the chargino mediated $b\to s$ transition is obviously more effective
in those cases where
the Higgsino mixing, $\mu$, is reduced. 
As we have already explained, this can be done with
non-universal Higgs masses where $\delta_2>0$ in
\eq{Higgsespara}. This is an interesting possibility, since a decrease
of $\mu$ also entails an increase of the Higgsino components of the
lightest neutralino and thus a larger detection cross section.
Although this would favour both cases a) and c) in \eq{nunivhiggs},
we have already seen in \fig{nunivsc50} how the B($\bmumu$) is more
stringent in the latter. 
This is due to the fact that in case c) the mass of the CP-odd Higgs
is also reduced. Thus for the same value of $\mu$, $m_A$ is smaller
and consequently B($\bmumu$) larger. In other words, case a) allows
lower values for the $\mu$ parameter and lighter CP-odd Higgses. Once
more, this leads to larger $\crosssec$.

\begin{figure}
  \epsfig{file=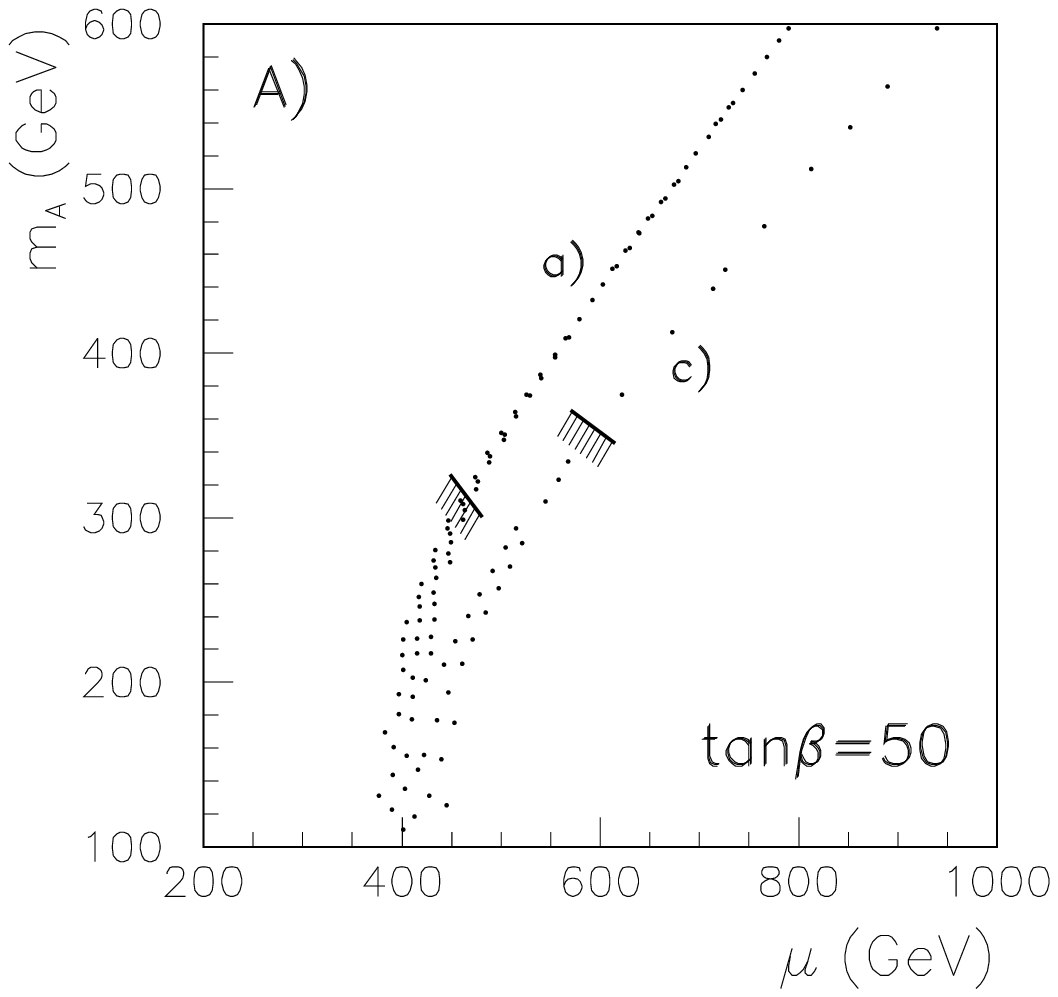,width=8cm}\hspace*{-0.5cm}
  \epsfig{file=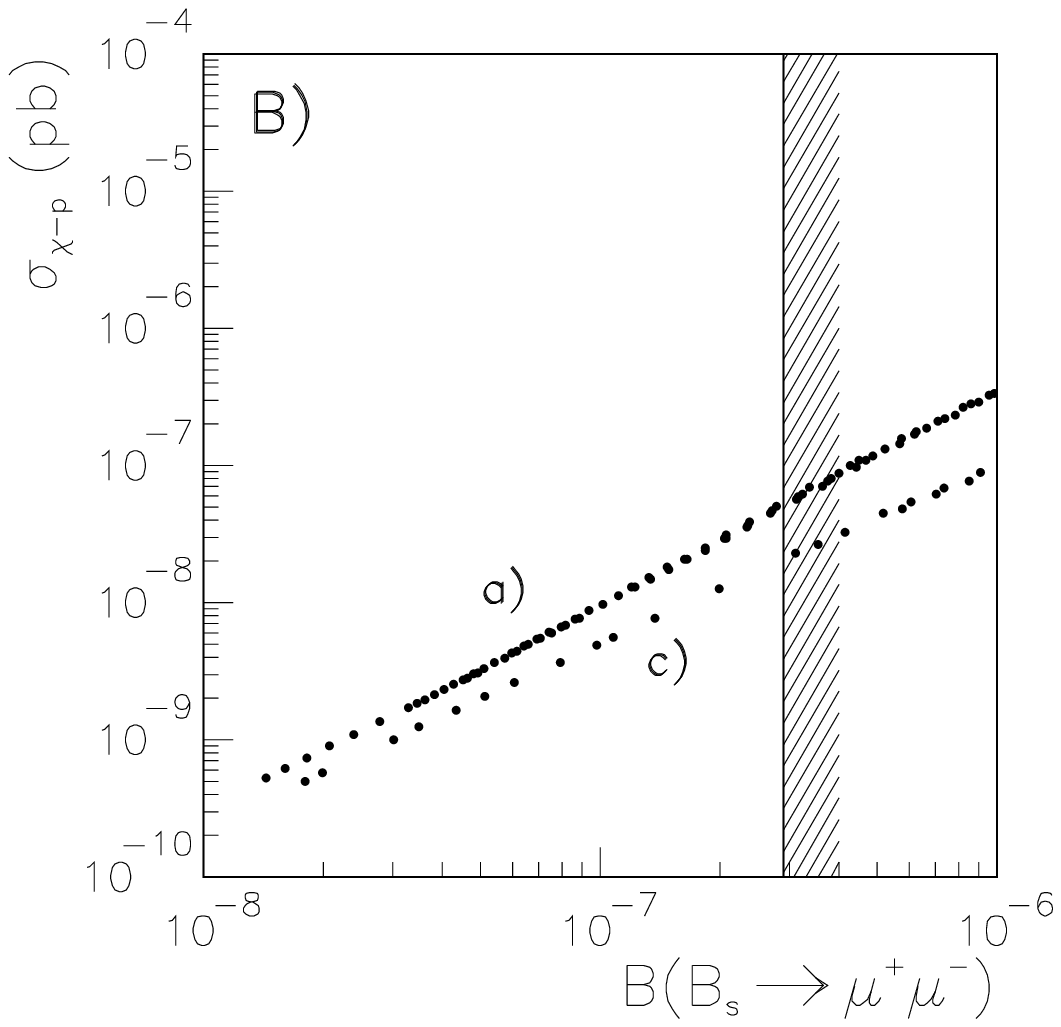,width=8cm}
    \captions{A) CP-odd mass versus the $\mu$ parameter for $\tan\beta=50$,
  $A=0$, and the two choices a) and c) in \eq{nunivhiggs} for
  non-universal 
  Higgses. All the points fulfil the experimental constraints and
  have the correct relic density. The constraint on B($\bmumu$) is
  displayed explicitly by means of a solid line with left shading
  which excludes the region below. B) Associated theoretical
  predictions for $\crosssec$ versus B($\bmumu$). The experimental
  limit on B($\bmumu$) is represented by the vertical line with right
  shading.}
  \label{cases}
\end{figure}

This is illustrated in \fig{cases}.  On the left-hand side 
we have plotted the value of the $\mu$ term versus the
CP-odd mass for cases a) and c), with $\tan\beta=50$ and $A=0$. 
Only those points fulfilling all
the experimental constraints and having the correct relic density are
displayed. Points below the solid line with left dashing
correspond to those excluded by the experimental constraint on
B($\bmumu$). Whereas in case c) the CP-odd mass and $\mu$ parameter
are 
bounded to
be $m_A\gsim 350$ GeV and $\mu\gsim600$ GeV, 
in case a) the value of B($\bmumu$)
remains below the experimental constraint for
$m_A\gsim 310$ GeV and $\mu\gsim475$ GeV. 
Consequently, $\crosssec$ is
slightly higher. The corresponding predictions for $\crosssec$ as a
function of B($\bmumu$) are represented on the right-hand side.
Larger values of
$\delta_2$ can help in obtaining even larger values for $\crosssec$
but obviously the decrease in $\mu$ is restricted by the experimental
bound on the chargino mass.

With this caveat in mind, 
let us now discuss the effect of adding non-universal gaugino
masses, which are known to provide a larger
flexibility in the neutralino sector \cite{cm04-1}. 
Due to the importance of the
gluino mass parameter,   
we will group the possible gaugino 
non-universalities in two different cases,
depending on whether the
ratio $M_3/M_1$ at the GUT scale 
decreases or increases with respect to its value in 
the universal
case. 
Interestingly, we will find how certain choices of non-universal
gauginos can help in decreasing B($\bmumu$) while allowing large values
for $\crosssec$.

\subsection{Decrease in $M_3/M_1$ ($\delta'_3<0$)}
\label{decm3m1}

When the ratio $M_3/M_1$ is decreased at the GUT scale (with
$\delta'_3<0$) heavier neutralinos, with a more important Higgsino
composition, can be obtained. This is due to the reduction of the $\mu$
parameter through the contribution of $M_3$ to the RGE of $\higgsu$. 
These neutralinos can have a
relatively high value for their direct detection cross section. 
Another consequence of the decrease in $M_3/M_1$
is the reduction in the value of the neutralino 
relic density, $\relic$. This may
be problematic, since the choices of non-universal scalars
\eq{nunivhiggs} 
already lead to a similar decrease, especially those where the Higgsino
components of $\neut$ increase.

Interestingly, the variation in the gaugino mass parameters also has
an influence on B($\bmumu$).
Namely, a decrease in $M_3$ leads to a smaller negative
contribution 
in the RGE of $A_t$, thus raising its value at the electroweak scale
and therefore
effectively reducing the stop mixing and increasing the stop mass.
This, together with 
the above-mentioned reduction of $\mu$, entails a decrease in
B($\bmumu$), as we already explained in the previous section.

\begin{figure}[!t]
  \hspace*{-1.5cm}\epsfig{file=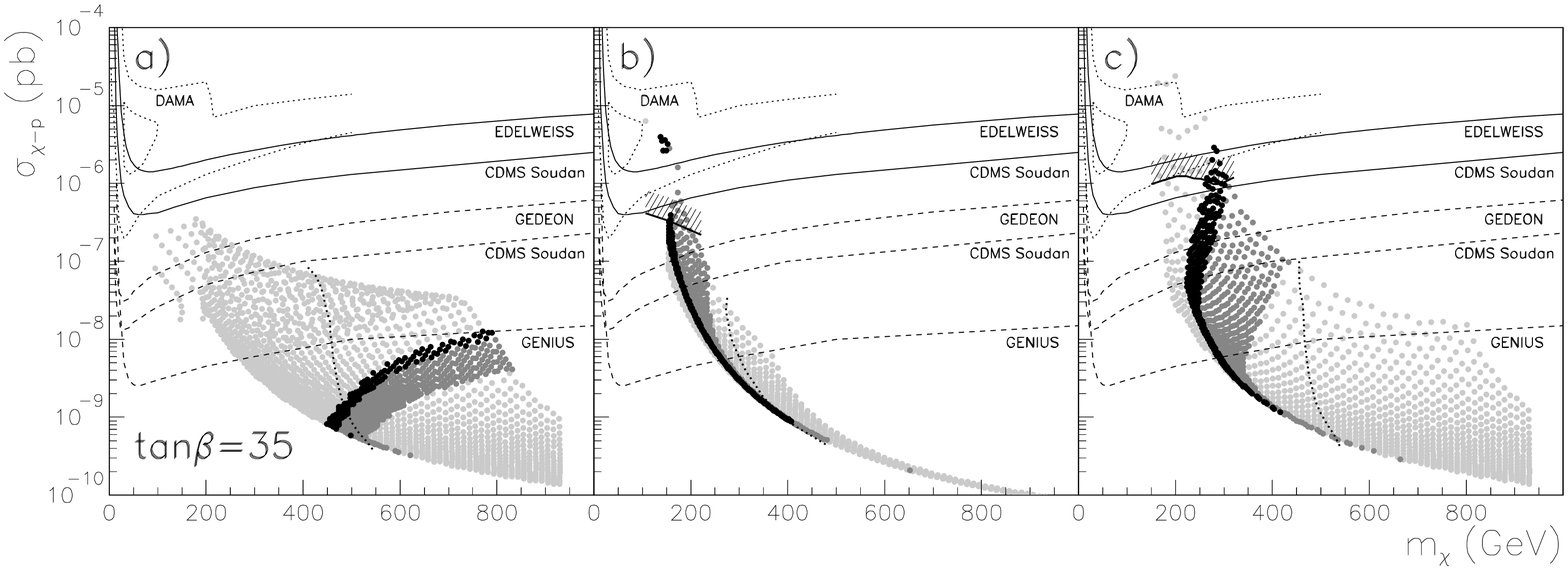,width=18cm}
  \captions{The same as \fig{nunivsc35} but for a case with
    $\delta'_{2,3}=-0.25$. }
  \label{cross35n}
  \hspace*{-1.5cm}\epsfig{file=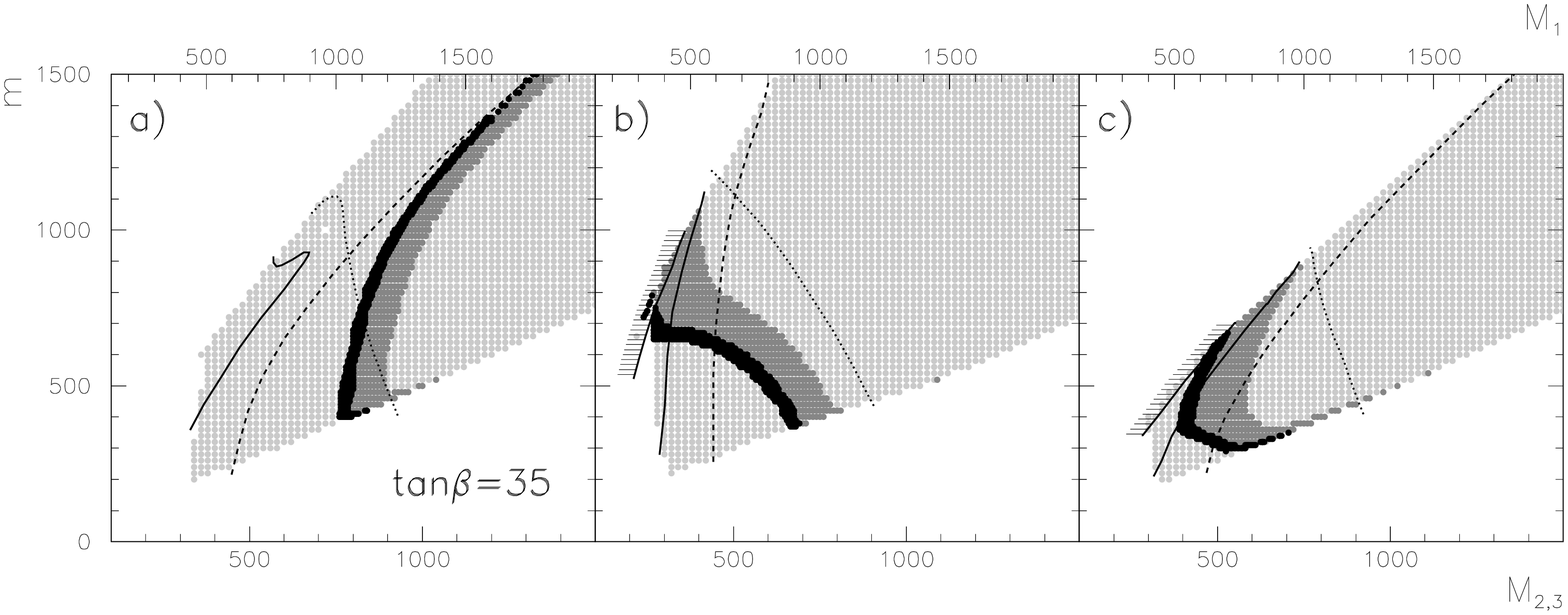,width=18cm}
  \captions{The same as \fig{nunivssparam35} but for a case with
    $\delta'_{2,3}=-0.25$.}
  \label{param35n}
\end{figure}

Let us begin by considering 
an example with $\delta'_{2,3}=-0.25$ in \eq{gauginospara}.
In \fig{cross35n} the neutralino-proton cross section is
plotted 
versus the neutralino mass, for $\tanb=35$, $A=0$, and a
scan in $m$ and $M$ ($m<1500$ GeV, and such that $50$ GeV
$<M_3<1500$~GeV)  
for the different choices of non-universal
scalar parameters \eq{nunivhiggs}. 
Notice that, as a consequence of the 
decrease in the
stop mixing and $\mu$ term,
the experimental constraint in B($\bmumu$) is not as stringent as in
the example with universal gauginos represented in \fig{nunivsc35}. 
In particular, points very close or even  
within the present sensitivities of present dark
matter detectors are obtained for cases b) and c), where
$\crosssec\lsim10^{-6}$ pb. In case a) the predictions for $\crosssec$
are smaller due to the less effective reduction of the CP-odd Higgs
mass.  

The corresponding $(m,M_i)$ parameter space is displayed in
\fig{param35n}. Notice that all the experimentally viable points in
case a) lie in a region which cannot be tested even by
GENIUS. Contrariwise, in
cases b) and c), CDMS Soudan and GENIUS will be able to test the
regions with $M_1\lsim500$ GeV and $M_1\lsim700$ GeV, respectively.

\begin{figure}[!t]
  \hspace*{-1.5cm}\epsfig{file=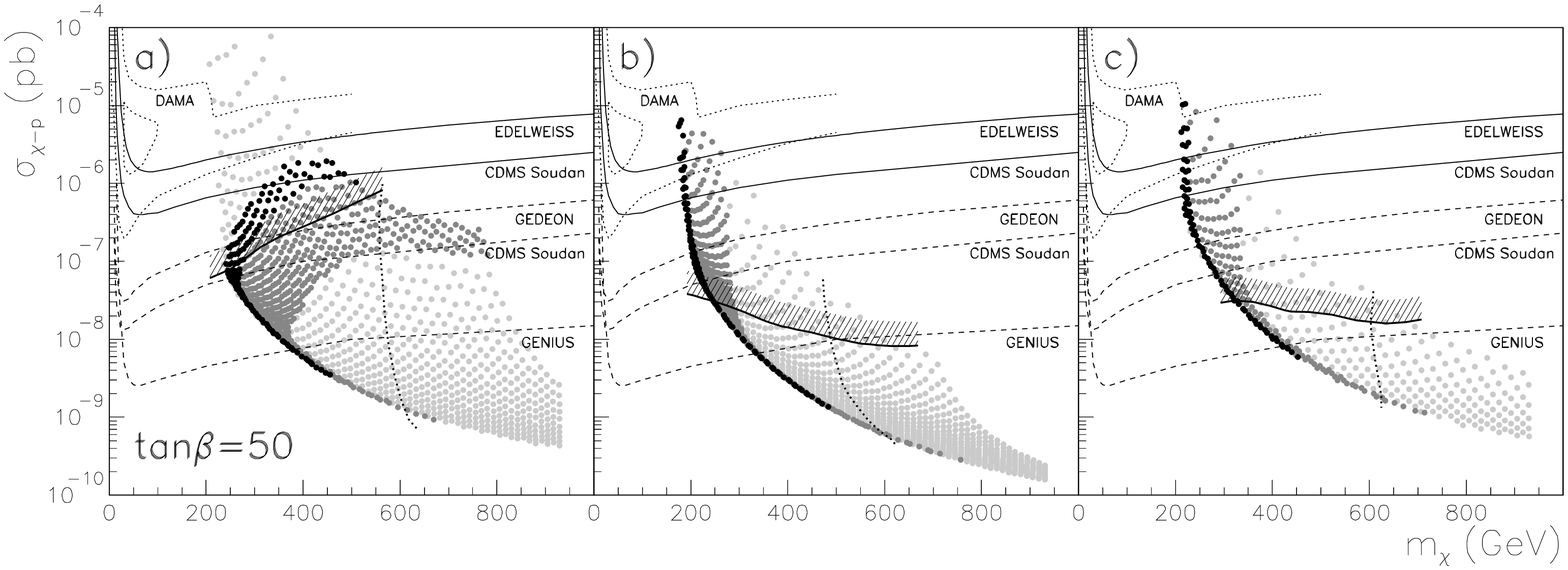,width=18cm}
  \captions{The same as \fig{nunivsc35} but for a case with
    $\tan\beta=50$ and 
    $\delta'_{2,3}=-0.25$.}
  \label{cross50n}
  \hspace*{-1.5cm}\epsfig{file=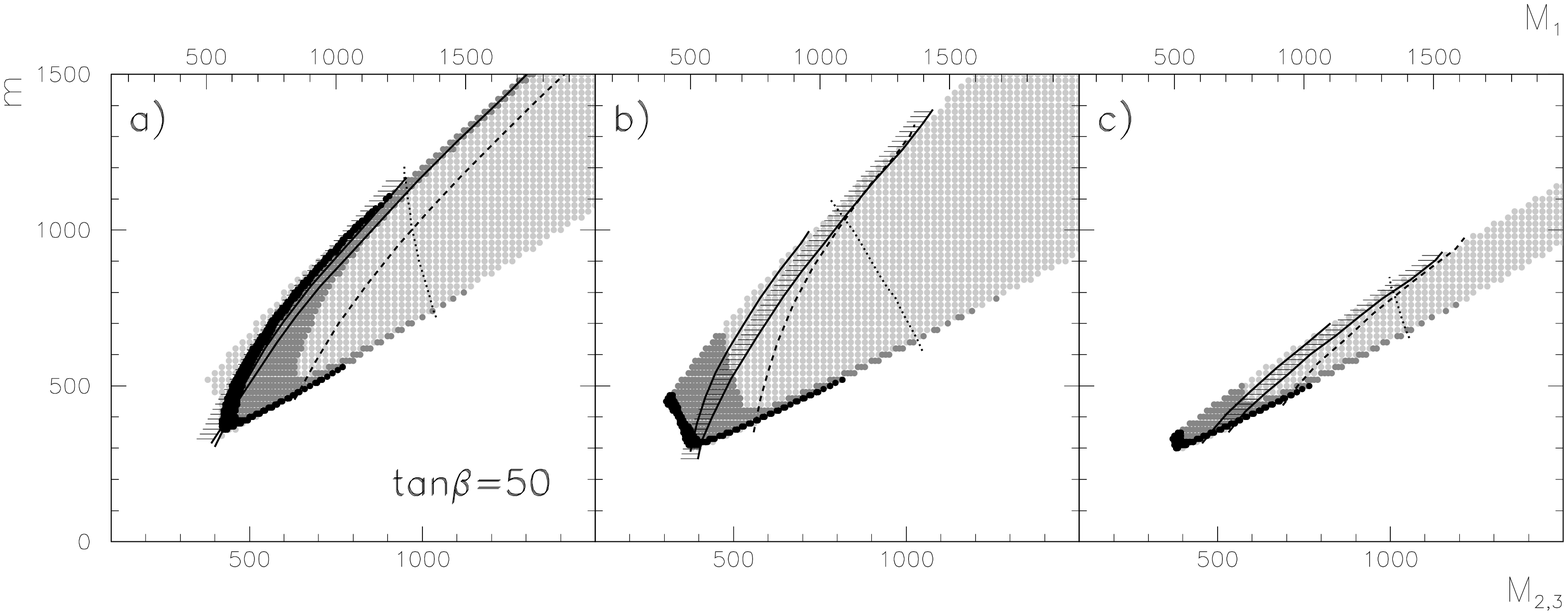,width=18cm}
  \captions{The same as \fig{nunivssparam35} but for a case with
    $\tan\beta=50$ and 
    $\delta'_{2,3}=-0.25$.}
  \label{param50n}
\end{figure}

A further example, now with $\tan\beta=50$, is represented in
Fig.\,\ref{cross50n}. 
Notice again how the B($\bmumu$) constraint is slightly less stringent
than in the examples with universal gauginos of \fig{nunivsc50}.
For instance, points appear in case a) which could be tested in the
future by the CDMS Soudan experiment with $\neumass\approx250$ GeV. 
In cases b) and c) the detection cross section 
is bounded at $\crosssec\lsim3\times10^{-8}$ pb, 
all the points lying below
the predicted sensitivity of 
CDMS Soudan.

The related $(m,M_i)$ parameter space is represented in
Fig.\,\ref{param50n}. We find that the only experimentally 
allowed regions in the parameter space where the WMAP result for the
relic density is obtained correspond to those in 
the coannihilation tail, close
to the line where the stau becomes the LSP. 
As expected, 
the constraint on B($\bmumu$) is responsible for the exclusion of
those areas where the CP-odd Higgs mass is smaller.

\begin{figure}
  \begin{center}
  \epsfig{file=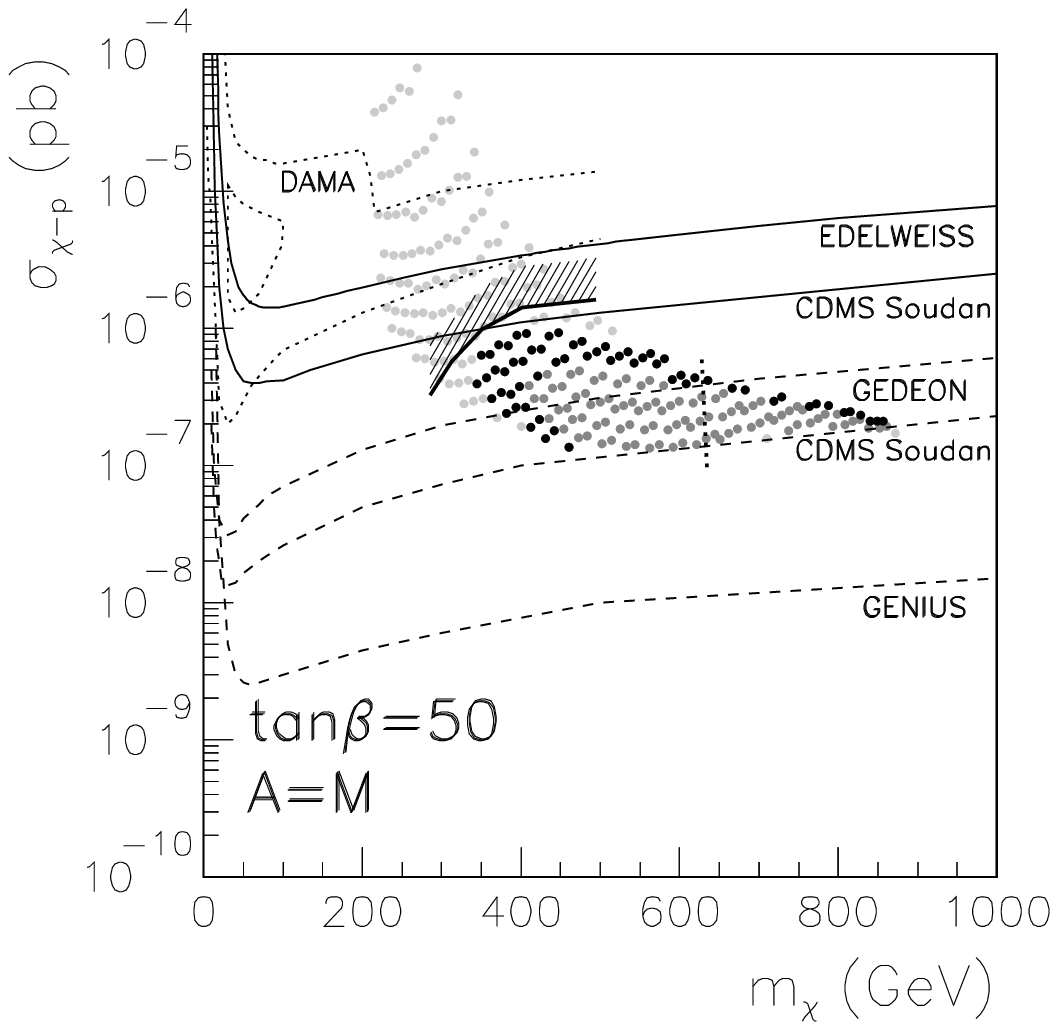,width=8cm}
  \end{center}
  \captions{The same as \fig{nunivsc35} but for a case with
  $\tan\beta=50$, $A=M$, non-universal Higgs masses with
  $\delta_1=0$ and $\delta_2=1.5$, and non-universal gauginos with
  $\delta'_{2,3}=-0.25$.} 
  \label{cross50dn}
\end{figure}

Once more, the B($\bmumu$) constraint can be softened and large
values for the neutralino direct detection cross section can be obtained 
by an appropriate choice of the initial parameters. As we commented in
the previous section, a modification in the trilinear parameter such
that it
minimises the stop mixing and a small $\mu$ term can have this effect.
Indeed, this possibility is now favoured by the fact that the decrease
of $M_3$ contributes to both effects.
To illustrate this we show in \fig{cross50dn} an example
with $\tan\beta=50$ and  non-universal gaugino masses with
$\delta'_{2,3}=-0.25$, where we have taken
$A=M$, and non-universal Higgs masses with $\delta_1=0$ and
$\delta_2=1.5$ in \eq{Higgsespara}.
We find how neutralinos fulfilling all experimental and astrophysical
constraints appear with $\crosssec\lsim10^{-6}$ pb and
$\neumass=350-400$ GeV. These neutralinos have a relatively large
Higgsino composition and are very close to the present
sensitivity of the CDMS Soudan detector.

\subsection{Increase in $M_3/M_1$ ($\delta'_3>0$)}
\label{incm3m1}

Increasing the value
of $M_3$ with respect to $M_1$ at the GUT scale can be done with
$\delta'_3>0$ in \eq{gauginospara}.
In this case, 
the constraint on the Higgs mass and on \bsg\ 
will be satisfied for smaller
values of $M$, and for this reason the effective value of $M_1$ can be
smaller than in the universal case.
Thus lighter neutralinos with a larger bino composition can be
obtained. 

Notice that in this case, being $M_3$ larger than in the universal
case, it contributes more to the running of $A_t$. This means that the
stop mixing is larger. Owing to this and to the slight increase in the
$\mu$ term, we can expect larger values for B($\bmumu$) and hence, 
a more stringent constraint.

\begin{figure}[!p]
  \hspace*{-1.5cm}\epsfig{file=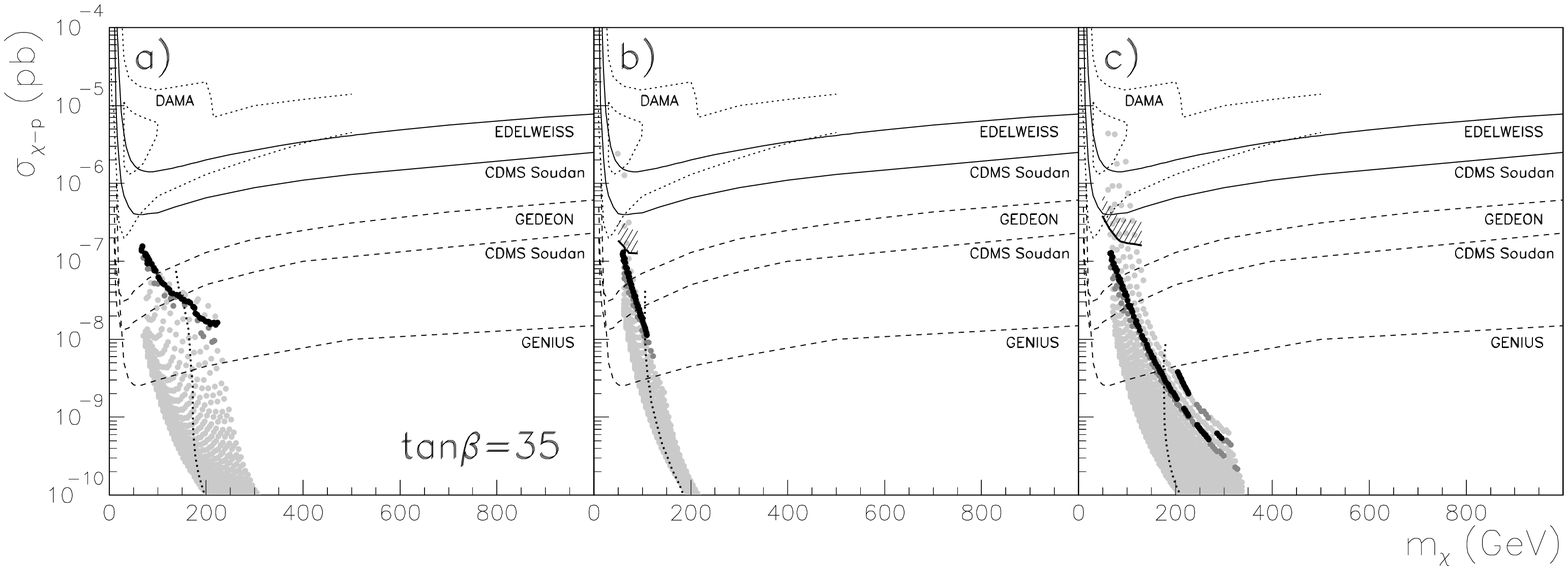,width=18cm}
  \captions{The same as \fig{nunivsc35} but for a case with
    $\delta'_{2,3}=1$.}
  \label{cross35w}
  \hspace*{-1.5cm}\epsfig{file=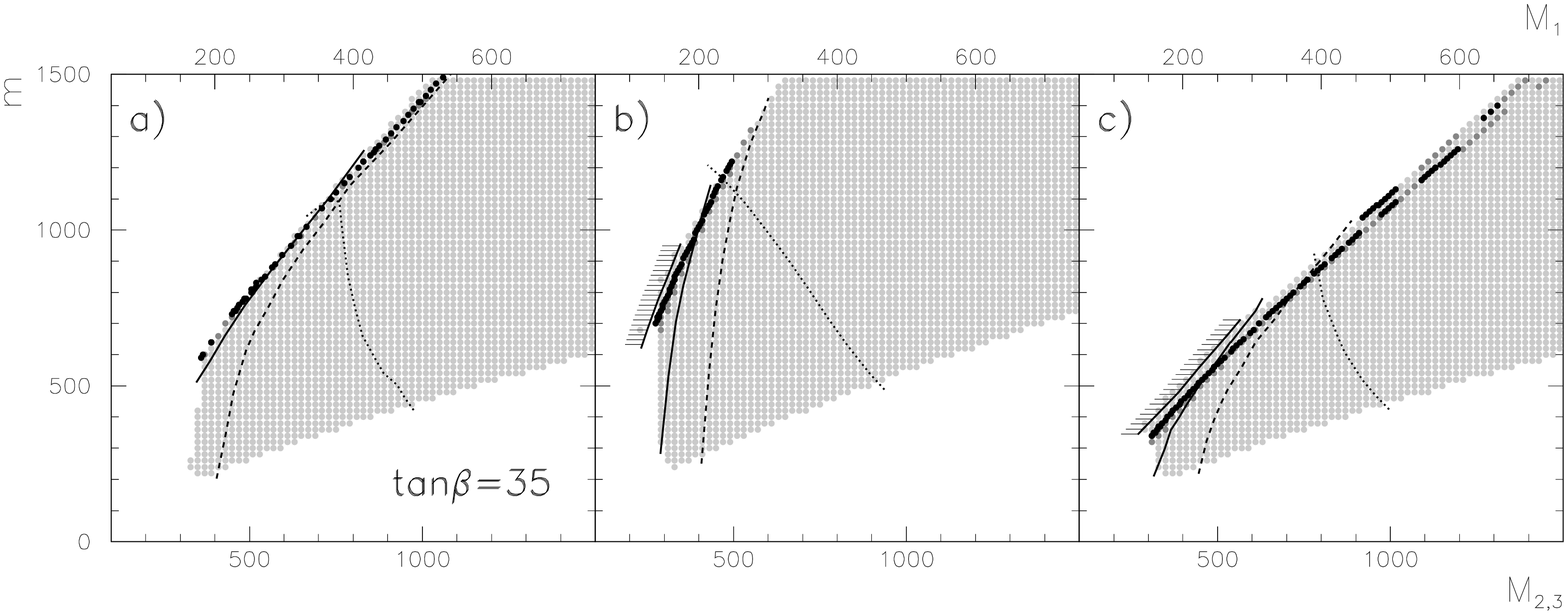,width=18cm}
  \captions{The same as \fig{nunivssparam35} but for a case with
    $\delta'_{2,3}=1$.}
  \label{param35w}
\end{figure}

\begin{figure}[!p]
  \hspace*{-1.5cm}\epsfig{file=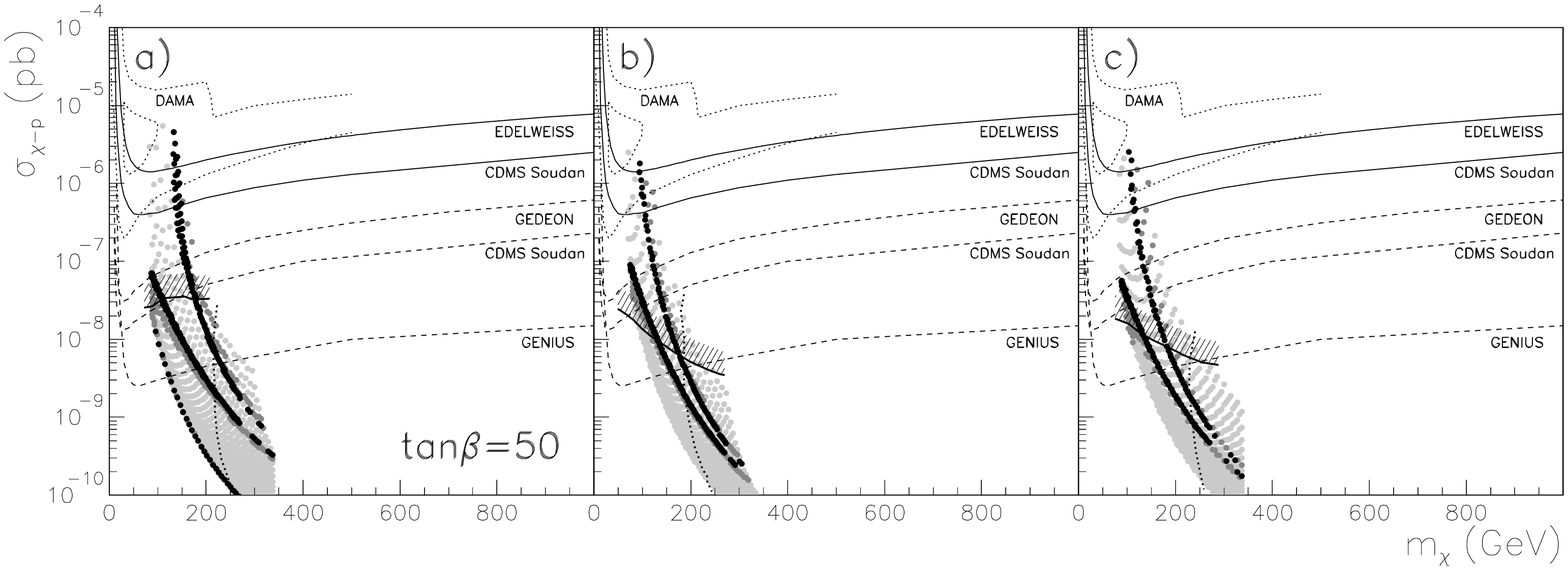,width=18cm}
  \captions{The same as \fig{nunivsc35} but for a case with
    $\tan\beta=50$ and 
    $\delta'_{2,3}=1$.}
  \label{cross50w}
  \hspace*{-1.5cm}\epsfig{file=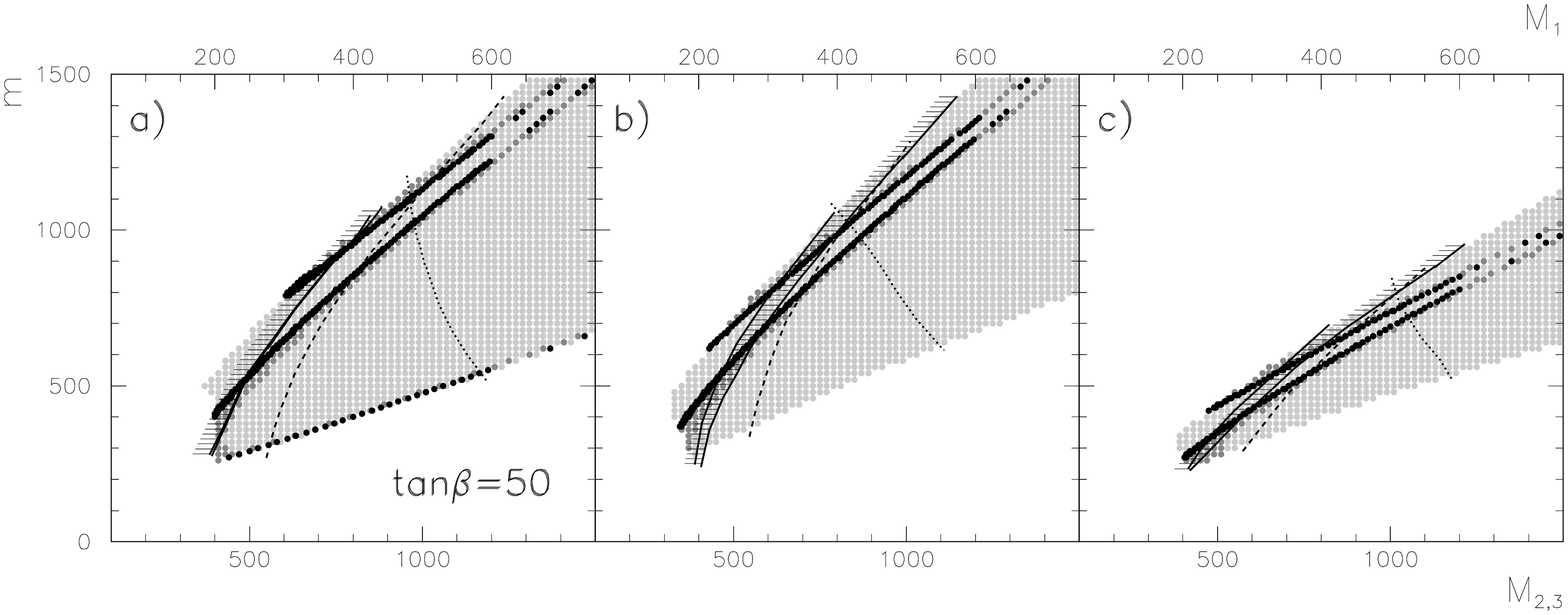,width=18cm}
  \captions{The same as \fig{nunivssparam35} but for a case with
    $\tan\beta=50$ and 
    $\delta'_{2,3}=1$.}
  \label{param50w}
\end{figure}

The theoretical predictions for $\crosssec$ as a function of the
neutralino mass are represented in \fig{cross35w} for an example with
$\tan\beta=35$, $A=0$, the three choices for non-universal Higgses
\eq{nunivhiggs},
and non-universal gaugino masses with 
$\delta'_{2,3}=1$ in \eq{gauginospara}.
As expected, lighter neutralinos than in those cases with
$\delta'_3<0$ shown in the
previous section, are obtained.
In the three cases the detection cross section is limited to be
$\crosssec\lsim10^{-7}$ pb, with neutralinos 
as light as $\neumass\gsim70$ GeV. 
These fall within the projected sensitivities of GEDEON and
CDMS Soudan. The experimental constraint on B$(\bmumu)$ plays no role
in limiting the value of $\crosssec$ of those points with the correct
relic density.

The associated $(m,M_i)$ parameter space is depicted in
\fig{param35w}. The absence of an allowed region in
the coannihilation tail in the three cases is remarkable. 
This is due to the occurrence of UFB
directions in the Higgs potential in those areas with light stops. 
Because of
the increase of $M_3$, the UFB constraints are more stringent and
the coannihilation regions tend to become excluded \cite{cm04-1}.
Regarding the B$(\bmumu)$ constraint, we find that the forbidden areas
in cases b) and c) only lie in regions where the neutralino relic density
is too low to reproduce the WMAP result.
Concerning the potential reach of future dark matter detectors, the 
whole allowed area in case a) could be
tested by GENIUS, while only those points with $M_1\lsim350$ GeV
would be within
the projected CDMS Soudan sensitivity. In cases b) and c) the area
covered by CDMS Soudan corresponds to $M_1\lsim200$ GeV and $250$ GeV,
respectively. In case c) GENIUS could test points up to $M_1\lsim375$
GeV.

Although for larger values of $\tan\beta$ one could expect a rise
in $\crosssec$, the associated increment in B$(\bmumu)$ renders this
possibility very constrained. For instance, the predictions for the
neutralino detection cross section are represented in \fig{cross50w}
for the former example but taking $\tan\beta=50$. Although points with
very large $\crosssec$ appear, 
all of these have a too large B$(\bmumu)$ and are ruled out. In the
end, $\crosssec\lsim2\times10^{-8}$ pb in case a) and
$\crosssec\lsim10^{-8}$ pb in cases b) and c), beyond the reach of the
projected CDMS
Soudan detector. 
Notice that these limits are slightly lower than in the case
represented in \fig{cross50n}, where $\delta'_3<0$. As we commented
above, this is due to the influence of $M_3$ in the stop mixing and
$\mu$ term.

\begin{figure}[!t]
  \hspace*{-1.5cm}\epsfig{file=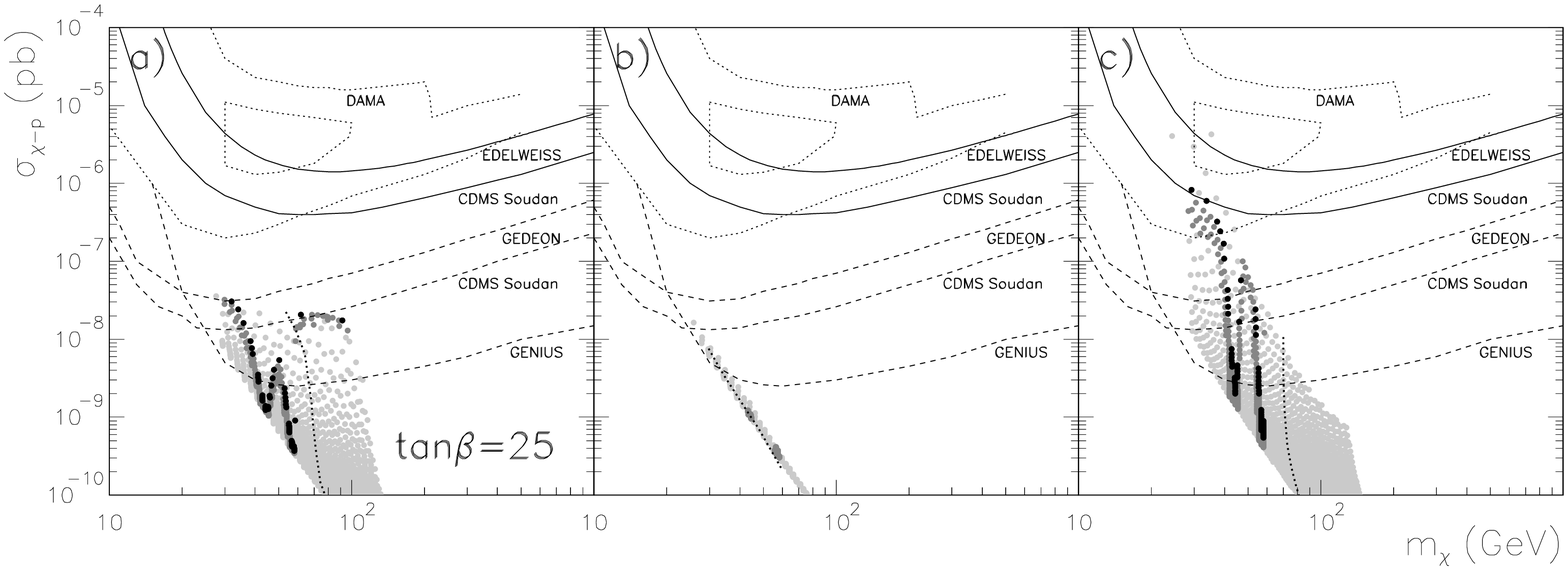,width=18cm}
  \captions{The same as \fig{nunivsc35} but for a case with
    $\tan\beta=25$ and 
    $\delta'_{2,3}=3$.}
  \label{cross25s}
  \hspace*{-1.5cm}\epsfig{file=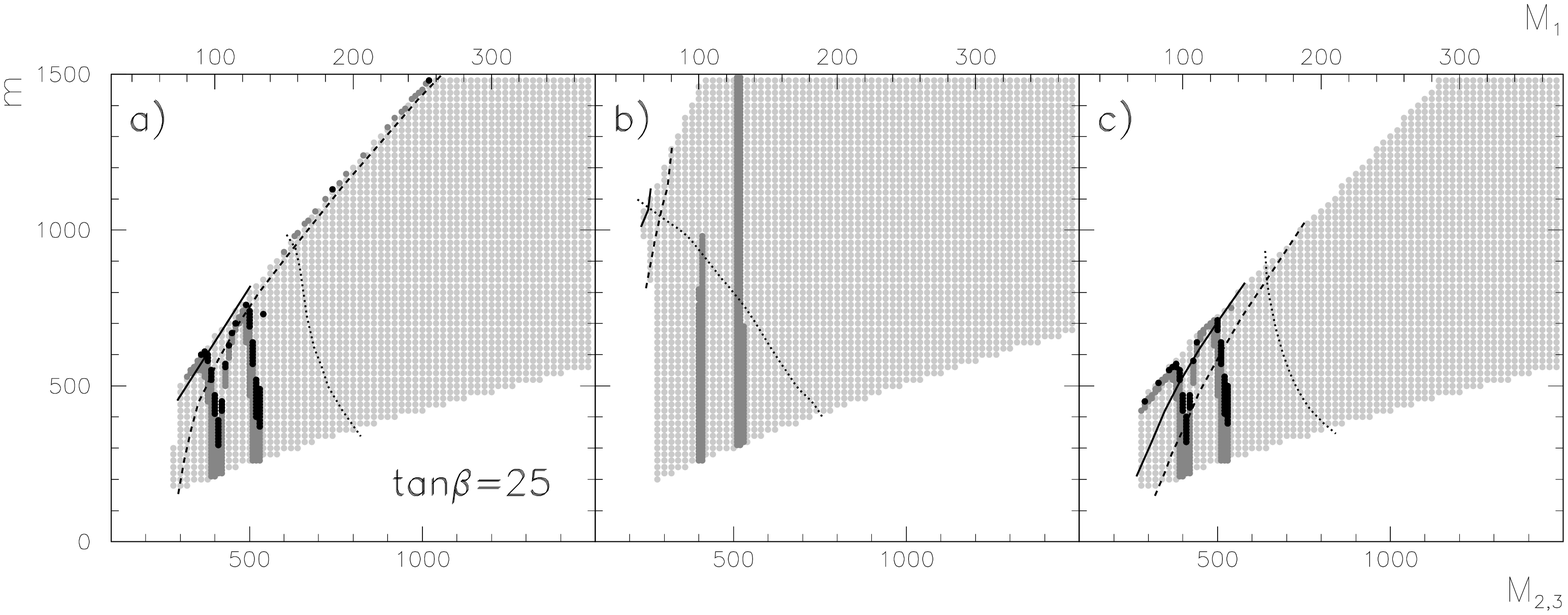,width=18cm}
  \captions{The same as \fig{nunivssparam35} but for a case with
    $\tan\beta=25$ and 
    $\delta'_{2,3}=3$.}
  \label{param25s}
\end{figure}

Only GENIUS would be able to test those regions with
$\neumass\lsim250$ GeV in case a) and $\neumass\lsim200$ GeV in b) and
c). As we see in \fig{param50w}, where the associated $(m,M_i)$
parameter is plotted, 
all the allowed points correspond to the regions where, due to
resonant annihilation of neutralinos mediated by the CP-odd Higgs, 
the correct relic density is obtained. This happens when
$2\neumass\approx m_A$.
Only case a) presents some allowed points in the coannihilation tail.

A further reduction in the neutralino mass is possible if larger
values are taken for $\delta'_{2,3}$. Obviously, the bino composition
of these neutralinos also increases. Figure \ref{cross25s} illustrates
this possibility for an example with $\delta'_{2,3}=3$,
$\tan\beta=25$, and $A=0$. Neutralinos as light as $\neumass\approx30$
GeV appear with large values for the detection cross section,
$\crosssec\approx10^{-6}$ pb. Interestingly, despite the lightness of
the Higgs masses that is necessary, due to the moderate values of
$\tan\beta$ these points are not in contradiction with the
experimental bound on B$(\bmumu)$.

\begin{figure}[!t]
  \hspace*{-1.5cm}\epsfig{file=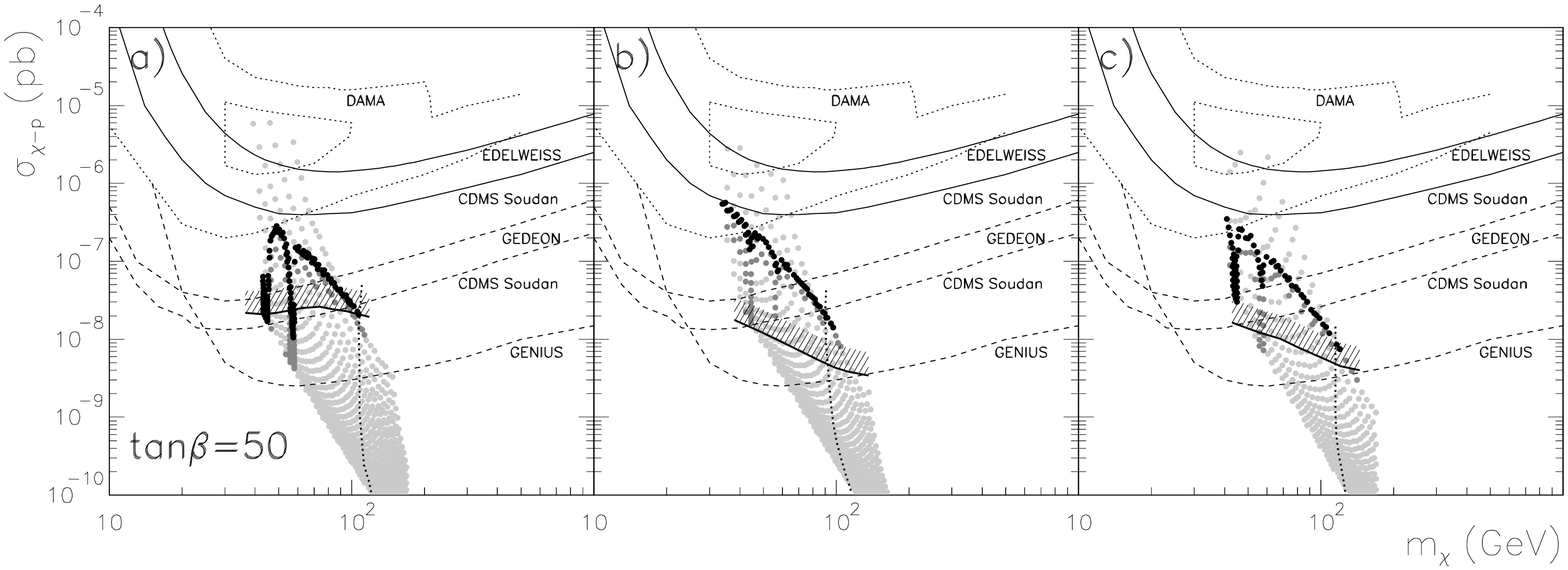,width=18cm}
  \captions{The same as \fig{nunivsc35} but for a case with
    $\tan\beta=50$ and 
    $\delta'_{2,3}=3$.}
  \label{cross50s}
  \hspace*{-1.5cm}\epsfig{file=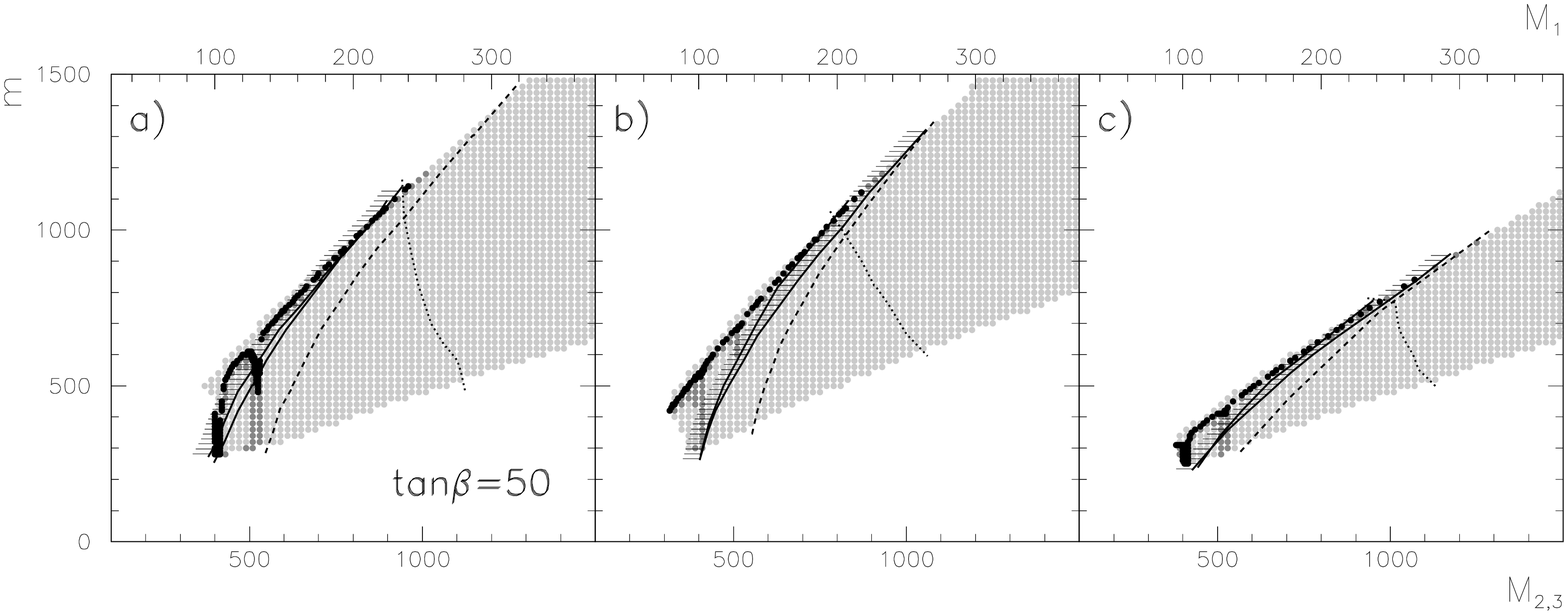,width=18cm}
  \captions{The same as \fig{nunivssparam35} but for a case with
    $\tan\beta=50$ and 
    $\delta'_{2,3}=3$.}
  \label{param50s}
\end{figure}

Nevertheless, these cases disappear as soon as $\tan\beta\gsim30$, in
which case the only regions of the parameter space for which all the
constraints are fulfilled coincide with the $s$-channel 
resonant annihilation of
neutralinos mediated by the lightest Higgs or the $Z$ boson, occurring
when $\neumass\approx{m_{h,Z}}/{2}$. For instance, the case with
$\tan\beta=50$ is represented in Fig.\,\ref{cross50s}.
The corresponding $(m,M_i)$ parameter space is shown in
Fig.\,\ref{param50s}. It can be seen how the combination of the
UFB and B$(\bmumu)$ constraints leave only the extremely fine-tuned
chimneys corresponding to resonant neutralino annihilation.

\subsubsection{Very light neutralinos}
\label{verylightneut}

Very light neutralinos can be obtained with non-universal gaugino masses
when $M_1\ll M_2,\,M_3,\,\mu$ (thus $\neut$ being bino-like) and when the
CP-odd Higgs is light enough ($m_A\lsim200$ GeV) in order to enhance
neutralino annihilation and obtain a relic density compatible with
WMAP. This can be achieved with the choices of Higgs
non-universalities we presented in \eq{nunivhiggs} for
$\tan\beta\gsim35$. 
Neutralinos obtained in this way can be as light as $10$ GeV
\cite{cm04-1}, and have a cross section within the reach of future
dark matter experiments.

However, due to the smallness of the Higgs masses and large values of
$\tan\beta$, these regions of the parameter space are likely to imply
sizable B$(\bmumu)$, and thus be very constrained by its improved 
experimental result.

\begin{figure}[!t]
  \hspace*{-1.5cm}\epsfig{file=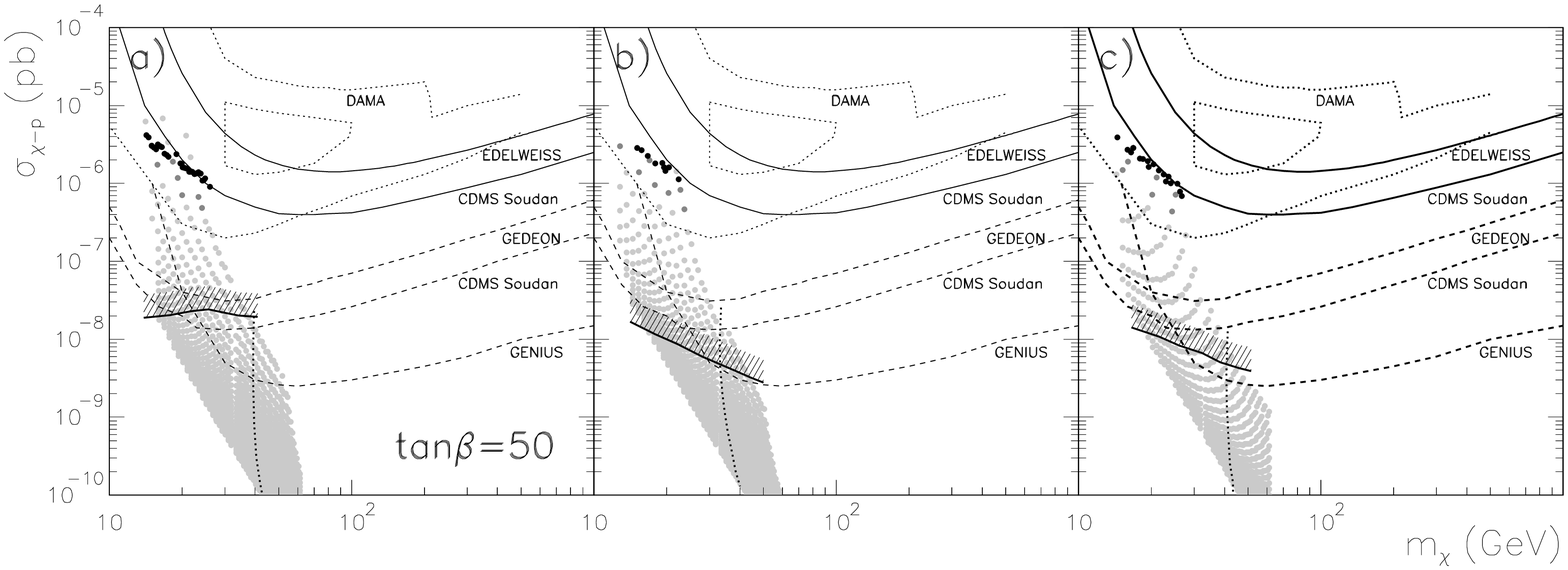,width=18cm}
  \captions{The same as in Fig.\,\ref{nunivsc35}, but with
    non-universal gauginos with, $\delta'_{2,3}=10$, and
    $\tan\beta=50$. 
  }
  \label{cross50q}
  \hspace*{-1.5cm}\epsfig{file=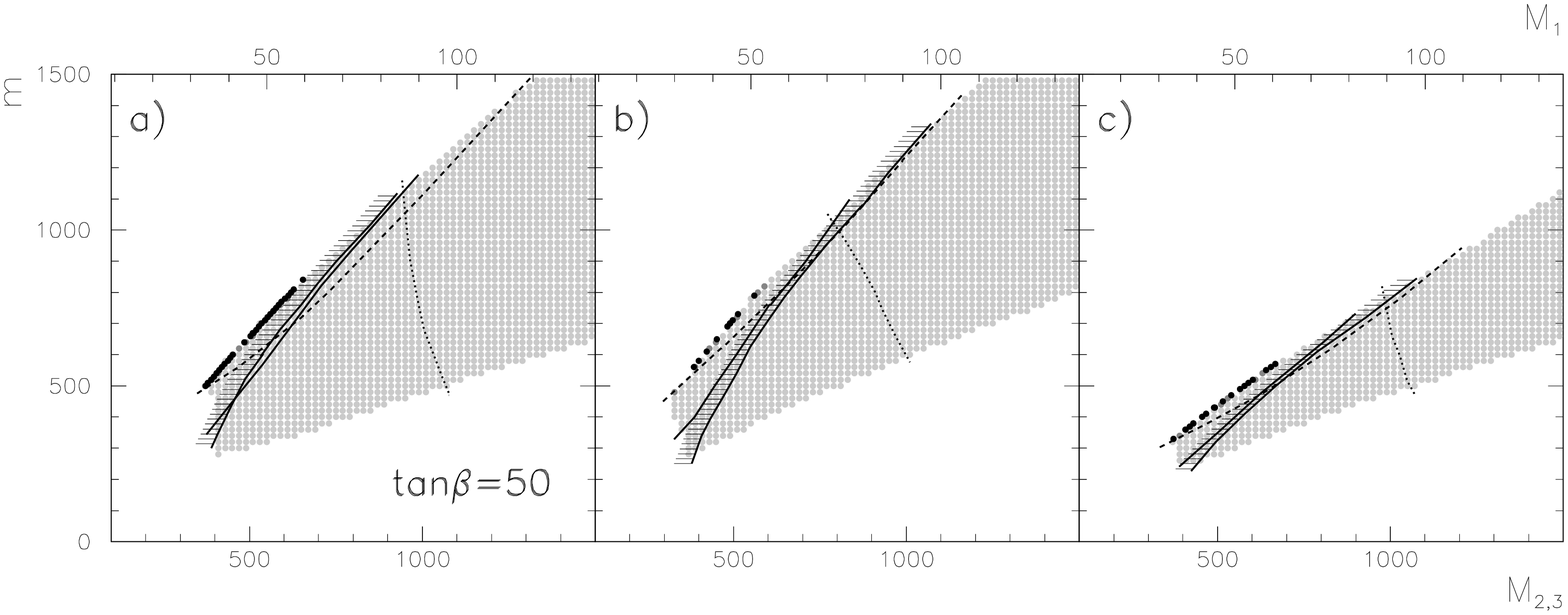,width=18cm}
   \captions{The same as \fig{nunivssparam35} but for a case with
    $\tan\beta=50$ and 
    $\delta'_{2,3}=10$.}
        \label{param50q}
\end{figure}

As an example, we show in Fig.\,\ref{cross50q} the case with
$\delta'_{2,3}=10$ and $\tan\beta=50$ for the three different choices
of Higgs masses \eq{nunivhiggs}. Neutralinos fulfilling the
experimental constraints and with a relic density in
agreement with WMAP appear 
with $\neumass\gsim15$ GeV and with very large values for the cross
section, $\crosssec\gsim10^{-6}$ pb. 
Unfortunately, once the experimental constraint on
B$(\bmumu)$ is applied, all these
cases are ruled out.
This is also shown in Fig.\,\ref{param50q}, where the $(m,M_i)$
parameter space is depicted. All the points that have the correct
relic density appear in a narrow line which runs alongside the region
where the CP-odd Higgs mass becomes tachyonic. This is contained
within the area excluded by the stringent B$(\bmumu)$ constraint.

Although in principle it could be 
possible to escape the drastic limitations imposed
by B$(\bmumu)$ by choosing suitable input parameters (which
lead to a decrease of 
the stop mixing and $\mu$ parameter), the amount of fine
tuning which is needed is huge. Besides, as already mentioned,
the CP-odd Higgs has to
remain very light to reproduce the correct value for the
relic density.
As a consequence, in order to induce a sufficient
decrease on the ratio $(\mu A_t/m_{\tilde t_L}^2)^2$ one needs such a
small value of 
$\mu$ that the constraint on the chargino mass and on \bsg\ become
difficult to fulfil. Also decreasing $ A_t/m_{\tilde t_L}$ is problematic since
it entails a decrease of the Higgs mass and
we are already very close to its experimental limit.

\section{Conclusions}
\label{conclusions}

In this paper we have performed an analysis of the direct detection
of neutralino dark matter in general supergravity scenarios, where
non-universality of soft scalar and gaugino masses can be present,
in the light of the recent experimental constraint on B$(\bmumu)$ from 
CDF and D0 Collaborations at the Tevatron.
More specifically, we have computed the theoretical predictions for
the spin-independent
neutralino-nucleon cross section, $\crosssec$, and compared it with the
sensitivities of present and future dark matter detectors.

We find that the B$(\bmumu)$ puts very strong limitations on 
$\crosssec$
in many interesting cases
of non-universal scalar and gaugino masses. 
Both observables are enhanced in the large $\tan\beta$ limit due to the
neutral Higgs exchange diagrams. 
Therefore, although the neutralino detection
cross section could be enhanced up to $\crosssec\sim 10^{-5}$~pb, 
those regions of the parameter space where the DAMA 
and the CDMS experiments could
detect signals are mostly excluded once the B($\bmumu$) constraint is
taken into account.
As a consequence, 
the resulting cross section typically becomes quite small, at the level
of  
$10^{-7}$ pb or even less.

Nevertheless, it is not impossible to evade this constraint and obtain
large neutralino detection cross sections. The correlation between
B($\bmumu$) and $\crosssec$ 
can be diluted if the top trilinear parameter is
tuned at the GUT scale and/or the $\mu$ parameter decreased 
in order to reduce the stop mixing and induce a smaller chargino
mediated $b\rightarrow s$ transition.

Some choices of scalar and gaugino non-universalities favour these
effects. In particular, we have shown that when the gluino mass
parameter, $M_3$, decreases with respect to the bino mass term, $M_1$,
heavier neutralinos with a larger Higgsino composition can be
obtained. In this case the B($\bmumu$) constraint is softer and, when
Higgs non-universalities are also considered,
$\crosssec\gsim10^{-6}$ pb can be obtained.

On the other hand, light bino-like neutralinos, which appear when $M_3$
increases with respect to $M_1$ are more difficult to reconcile with
the experimental constraint on B($\bmumu$). Still, neutralinos as
light as $\neumass\gsim30$ GeV and with $\crosssec\lsim 10^6$ pb, within
the DAMA region, can be
obtained with moderate values of $\tan\beta$.

\vspace*{1cm}
\noindent{\bf Acknowledgements}

The work of SB was supported by NSERC of Canada.
DGC gratefully acknowledges financial support from the 
PPARC and from the 
European Network
for Theoretical Astroparticle Physics (ENTApP), member of ILIAS, EC
contract number RII-CT-2004-506222.
YGK was supported by the Korean Federation of Science and
Technology Societies through the Brain Pool program. 
PK was supported in part by KOSEF Sundo grant R02-2003-000-10085-0, and
KOSEF through CHEP at Kyungpook National  University. 
The work of CM was supported in part by the Spanish DGI del MEC
under Acci\'on Integrada Hispano-Alemana HA2002-0117, and under
Proyectos Nacionales BFM2003-01266 and FPA2003-04597,
and also by the European Union under the RTN program MRTN-CT-2004-503369, and
under the ENTApP Network of the ILIAS project RII3-CT-2004-506222.

DGC would like to thank A. Dedes for useful discussions.

\providecommand{\href}[2]{#2}

\end{document}